\documentclass{article}
\PassOptionsToPackage{numbers, compress}{natbib}
\usepackage[table]{xcolor}  
\usepackage{longtable} 
\usepackage{rotating}  

\usepackage{tabularx} 
\usepackage{changepage}
\usepackage{graphicx}
\usepackage{array}                
\usepackage[most]{tcolorbox} 

\usepackage[final]{neurips_2024}
\usepackage[utf8]{inputenc} 
\usepackage[T1]{fontenc}    
\usepackage{hyperref}       
\usepackage{url}            
\usepackage{booktabs}       
\usepackage{amsfonts}       
\usepackage{nicefrac}       
\usepackage{microtype}      
\usepackage{fontawesome}
\usepackage[table]{xcolor}  
\usepackage{caption}        

\newcolumntype{C}[1]{>{\centering\arraybackslash}p{#1}}

\usepackage{etoolbox}
\makeatletter
\patchcmd{\endabstract}{\vskip 1ex}{%
  \vskip 1ex%
  {\footnotetext{Authors listed alphabetically. Author contributions are detailed in the contribution statement.}}%
  {\footnotetext{\faEnvelopeO\ \texttt{sbarrett.work321@gmail.com}}}%
  {\footnotetext{The relevant code for Section 4 can be found at \url{https://github.com/atlaie/delphi-llms}}}%
}{}{}
\makeatother

\title{Assessing confidence in frontier AI safety cases}

\author{%
  Steve Barrett \faEnvelopeO \And Philip Fox \And Joshua Krook \AND Tuneer Mondal \And Simon Mylius \And Alejandro Tlaie \AND
  {\normalfont Arcadia Impact -- AI Governance Taskforce}
}

\begin{document}

\maketitle
\begin{abstract}
Powerful new frontier AI technologies are bringing many benefits to society but at the same time bring new risks. AI developers and regulators are therefore seeking ways to assure the safety of such systems, and one promising method under consideration is the use of safety cases. A safety case presents a structured argument in support of a top-level claim about a safety property of the system. Such top-level claims are often presented as a binary statement, for example “Deploying the AI system does not pose unacceptable risk”. However, in practice, it is often not possible to make such statements unequivocally. This raises the question of what level of confidence should be associated with a top-level claim. We adopt the Assurance 2.0 safety assurance methodology, and we ground our work by specific application of this methodology to a frontier AI inability argument that addresses the harm of cyber misuse. We find that numerical quantification of confidence is challenging, though the processes associated with generating such estimates can lead to improvements in the safety case. We introduce a method for better enabling reproducibility and transparency in probabilistic assessment of confidence in argument leaf nodes through a purely LLM-implemented Delphi method. We propose a method by which AI developers can prioritise, and thereby make their investigation of argument defeaters more efficient. Proposals are also made on how best to communicate confidence information to executive decision-makers. 
\end{abstract}

\newpage

\section*{Executive Summary}

\textbf{What does it mean to have confidence in a safety case and why is such confidence needed? (Sections 2, 3)}

\begin{itemize}
    \item Safety cases generally make binary claims such as “Deploying the AI system does not pose unacceptable cyber risk”. However, a decision-maker or evaluator is then left asking the question: what confidence is there that the claim is true? 
\item Part of the answer involves showing that the argumentation is logically sound and complete. Another part of the answer is to enable the degree of belief in a top-level claim to be expressed as a probability. Such probabilistic quantification of confidence would be valuable because it could help an AI developer to know what doubts about the safety case are of most concern, as well as to determine when sufficient safety assurance work has been done.
\item All these aspects are covered by a safety assurance framework called Assurance 2.0, and it is this framework which we have applied to the problem of establishing confidence in a frontier AI ‘inability’ safety argument for the case of cyber misuse. 
\end{itemize}
\textbf{How can confidence be assessed in a way that is as reproducible, transparent and as non-subjective as possible? (Section 4)}
\begin{itemize}
    \item It can be difficult for third parties to have high confidence in an argument produced by an AI developer where the probabilistic evaluations of the level of truth in a claim are not reproducible on account of being made subjectively or including reasoning which is not transparent.
\item The process of establishing and assessing confidence in safety cases can sometimes require assigning probabilities to possible future events or outcomes, which are often conditioned on existing relevant information or evidence. We describe a method for addressing this problem using a variation of the Delphi method, in which the role of the experts is performed by LLMs. This new approach for use in establishing probabilistic confidence in safety cases is reproducible and has the potential to offer more transparency compared to the situation where human experts provide probabilistic valuations. We believe the approach is worthy of further investigation and consideration. 
\end{itemize}

\textbf{How to identify defeaters (i.e. doubts and challenges about the safety case)? (Section 5)}
\begin{itemize}
    \item Given the inevitable blindspots of AI developers in assessing their own systems, a comprehensive search for defeaters requires critical review by both internal and external (third-party) challengers. We suggest following a “dialectical method”, wherein safety case challengers have the explicit goal of finding flaws in the AI developer’s reasoning and are thus well-placed to counter biases that may otherwise exist within the teams developing the AI system and its safety case.
\end{itemize}

\textbf{How can it be quantitatively determined that the level of confidence in a safety case has become sufficient such that the system can be deployed? (Section 7)}
\begin{itemize}
    \item Belief in the claims of an argument can be quantified probabilistically, for example by a variation of the Delphi method described above. These argument leaf-node probabilities can be aggregated (propagated up through the safety case) to obtain a quantified probabilistic assessment of confidence in the top-level claim.
\item We applied Assurance 2.0’s ‘sum of doubts’ and ‘product’ methods for probabilistic quantification to a small fragment of a cyber misuse safety case. 
\item We found that, for both methods, in order to achieve a modestly high $(95\%)$ confidence in the overall claim for this small safety case fragment (comprising just 7 argument components), we required very high $(~99.3\%)$ confidence in each of the components. 
\item We conjecture that it would be very challenging to achieve such high levels of confidence in the argument leaf nodes, in particular where such confidence assessments require judgements to be made about uncertain (e.g.) future outcomes. We would anticipate that the challenge of achieving high confidence will only get greater when considering the safety argument as a whole, rather than just a small fragment of it. 
\item It should be cautioned however, that our results for the cyber misuse safety case fragment were for a particular argument structure (logically conjunctive and independent claims), and it would be worth considering how things might change, and potentially improve in cases where alternative argument structures are possible, for example in cases where multiple diverse sub-arguments could be used to support a single claim. 
\item Things might also improve where a probabilistic target forms an intrinsic internal part of the argument itself. In summary, it remains an open question whether realistic and practical methods might be found to achieve the very high probabilistic confidences in safety cases that would be required for the most serious, or even catastrophic, harms. Our results certainly point to it being a non-trivial challenge, and one that will require further investigation. 
\end{itemize}

\textbf{How can AI developers prioritise which defeaters to tackle first? (Section 8)}
\begin{itemize}
    \item Defeaters may be eliminated through system and/or safety case modification. The order with which these doubts are tackled affects the efficiency with which an AI developer’s workforce can converge upon an acceptable system design and associated safety case. We describe a methodology that AI developers can follow, that takes into account the defeater’s potential impact on probabilistic confidence in the top-level claim, potential impact on the logical soundness of the argument, the probability of the defeater being sustained and the expected effort required to resolve the defeater.
\end{itemize}

\textbf{How can the level of confidence in a safety case best be communicated to executive decision-makers? (Section 9)}
\begin{itemize}
    \item Decisions on whether to deploy an AI or cease its operation will be made by executives who may have limited technical background.To the extent possible, we advocate for the use of visual as opposed to textual communication of confidence.
\end{itemize}

\textbf{Recommendations (Section 10)}

\begin{enumerate}
    \item As AI capabilities increase, leading to an increased risk of harm, it is recommended that confidence assessment of safety cases be performed using techniques like those provided by Assurance 2.0. Such techniques, ensure the systematic consideration of the many important dimensions to the challenge of confidence assessment, and despite the very significant difficulties in obtaining credible quantitative assessments of confidence, will lead to the production of higher quality safety cases.
    \item As far as we are aware, there are no standardised guidelines for what methods of confidence assessment should be used by developers of frontier AI. It would be useful to develop and provide such guidelines.
    \item For defeater identification to be comprehensive and credible, ‘safety-case red-teamers’ who are external to the AI developer should be involved.
\end{enumerate}

\newpage

\tableofcontents

\vspace{9cm}

\section*{Contribution Statement}

Author Contributions (CRediT taxonomy)

\textbf{Conceptualization:} Steve Barrett, Ben R. Smith \\
\textbf{Methodology, Investigation:} All authors \\

\textbf{Writing – Original Draft, Review:} \\
\vspace{0.2cm}
\begin{tabular}{@{}ll@{}}
$\quad$ Steve Barrett      & Sections 1, 2.1, 2.2, 3.2, 6, 10 \\
$\quad$ Philip Fox         & Sections 2.2, 3.1, 5, 5.1, 10 \\
$\quad$ Joshua Krook       & Sections 9, 10 \\
$\quad$ Tuneer Mondal      & Sections 7, 10 \\
$\quad$ Simon Mylius       & Sections 5.2, 8, 10 \\
$\quad$ Alejandro Tlaie    & Sections 4, 5.1, 6, 10 \\
\end{tabular}

\textbf{Writing – \LaTeX:} \\
\vspace{0.2cm}
$\quad$ Alejandro Tlaie \\

\textbf{Writing – Review \& Editing, Supervision:} \\
\vspace{0.2cm}
$\quad$ Steve Barrett (Lead Editor) \\

\textbf{Project Administration:} Steve Barrett, Ben R. Smith

\newpage

\section{Introduction}

‘Frontier AI’ refers to a class of the most advanced, highly capable, general-purpose AI models that can perform a wide variety of tasks, and which currently primarily encompass foundation models consisting of very large neural networks using transformer architectures. These powerful new frontier AI technologies can bring many benefits to society, but at the same time can bring new risks. Companies that are developing frontier AI systems are therefore seeking methods to assure the safety of their systems before and during deployment, and potentially even before commencing training. One promising method for assuring the safety of such systems is through the use of safety cases \cite{buhl2024safety}. 

Safety cases make use of a structured argument that is supported by evidence in order to make a claim about the safety properties of a system. Whilst many elements of a safety case may be very well supported, in a typical safety case, there will nevertheless remain some sources of uncertainty and residual doubt. These uncertainties may, by way of example, occur due to the inherent nature of the argument that is being made (e.g. including inductive components vs. purely deductive), due to the type and quality of the evidence provided or due to a multitude of other possible factors. Hence, when making a decision on whether or not to deploy an AI model, a decision maker can be expected to benefit from information detailing firstly what level of confidence should be placed in the assurance claim that is being made, and secondly the specifics of the main sources of uncertainty and residual doubt that influence this level of confidence.

This paper applies one of the foremost approaches for building a safety case and determining an associated level of confidence, Assurance 2.0 \footnote{A number of relevant papers, including the 2-pager “Assurance 2.0 in a nutshell”  have been collected by John Rushby, see \url{https://www.csl.sri.com/users/rushby/assurance2.0}} \cite{bloomfield2022assessing} to identify what particular lessons can be drawn when assessing confidence in safety cases of frontier AI systems. To illustrate how our findings can be applied in practice we show how the concepts can be applied with reference to the safety case template for an inability argument for the cybersecurity misuse harm that is described in \cite{goemans2024safety}. In an ‘inability’ argument a case is made that the AI system lacks the ability to cause harm. This is in contrast to other types of argument, for example a ‘control’ argument where the AI system does have the ability to cause harm, but is controlled so as not to do so, or a ‘trustworthiness’ argument where again the AI system has the ability to cause harm but a case is made that it can be trusted not to do so \cite{clymer2024safety}. We focus on the cyber misuse harm and the inability safety case, not only because of the availability of the safety case template but also because dealing with cyber misuse and creation of good inability safety cases represent near-term, and hence high priority challenges for the community. Cyber misuse is also a class of harm for which there is a relatively well-developed understanding of risk. 

In \textbf{Section \ref{sec2}} of the paper the rationale for selecting the Assurance 2.0 methodology is described and a high-level introduction to the technique is provided. \textbf{Section \ref{sec3}} describes how confidence in the logical validity and soundness of the safety case can be attained. \textbf{Section \ref{sec4}} describes how to assess probabilistic confidence in the leaf nodes of an argument. \textbf{Section \ref{sec5}} describes how confidence can be enhanced through processes that enable the case to be challenged such that sources of doubt can be surfaced. \textbf{Section \ref{sec6}} describes how to gain confidence when residual risks must remain in the safety case. \textbf{Section \ref{sec7}} addresses the question of how quantified assessments of confidence, made for the various elements comprising the argument, can be propagated up into an overall confidence assessment of the top-level claim of the safety case. \textbf{Section \ref{sec8}} addresses the issue of how to prioritise the resolution of defeaters (doubts about the safety case). \textbf{Section \ref{sec9}} identifies the preferred form that confidence level information should take for it to be communicated to, and be most actionable by executive decision-makers. Finally, the overall conclusions are provided in \textbf{Section \ref{sec10}}. 

\section{Methodology for determining confidence}\label{sec2}
In this section we explain the requirement for, and the benefits of, providing confidence measures. We then go on to explain our rationale for selecting the Assurance 2.0 methodology \cite{bloomfield2022assessing}, for the purposes of conducting our study into confidence assessment of frontier AI safety cases. 

\subsection{Requirement for a confidence measure}

\begin{figure}[htbp]
  \centering
  \includegraphics[width=1\textwidth]{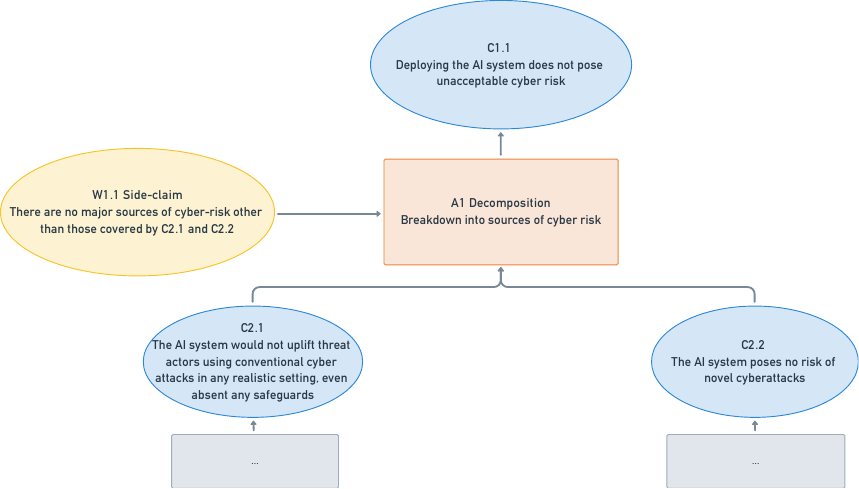}
  \caption{Top-level claim and its decomposition for the cyber misuse harm \cite{goemans2024safety}}
  \label{fig:top_claim}
\end{figure}

In the case of the cyber-misuse inability argument provided in \cite{goemans2024safety}, the top-level safety claim for the cyber-misuse harm is “\textit{C1.1: Deploying the AI system does not pose unacceptable cyber risk}”, and this is itself decomposed into two sub claims “\textit{C2.1: The AI system would not uplift threat actors using conventional cyberattacks in any realistic setting even absent safeguards}” and “\textit{C2.2: The AI system poses no risk of novel cyberattacks}” (Figure \ref{fig:top_claim}). For this cyber misuse inability case, and for most practical safety cases, such claims cannot be made unequivocally. The reader of a safety case is therefore often left asking the question: what confidence is there that the claims are true?

The level of confidence in the top-level claim of an argument will be a function of the level of confidence in the argument that supports the top-level claim and hence a function of the level of confidence in all the sub-claims, arguments, evidence, justifications, etc, which comprise that argument. The level of confidence in the top-level claim will also be a function of the processes used in constructing the argument, for example those used to ensure the logical validity of the argument and those used to counteract potential biases, all of which are intended to ensure that the overall argument is complete and sound. 

A developer or an evaluator of a safety case may find it useful to create, or be provided with, a quantified level of confidence in a top-level claim. Such quantification could be used in determining whether sufficient safety assurance work has been done, in quantitatively comparing the merits of different argument structures and types, or in better understanding and communicating deployment risk. Bloomfield and Rushby \cite{bloomfield2022assessing} describe confidence as the degree of belief in the top-level claim and state that if this confidence were to be quantified then it would most naturally be expressed as a probability. As such, a probabilistic confidence of 0 in the top-level claim would mean that the probability of the top-level claim being adequately supported by the argument would be 0, meaning that there is no confidence in the top-level claim, whilst in contrast, a probabilistic confidence of 1 would mean that there was complete confidence that the top-level claim is adequately supported by the argument.

The top-level claim that we are considering in this paper is “\textit{C1.1: Deploying the AI system does not pose unacceptable cyber risk}” (Figure \ref{fig:top_claim}). Risk is itself often defined as being some product of the severity of harm, and the probability of that harm being realised. The concept of making a probabilistic quantification of confidence in a claim which itself contains a probabilistic element can be confusing. An analogy may be useful. A weather forecaster may say that they expect that it will be sunny all day tomorrow, which is a probabilistic claim about some future state of the world (analogous to a claim/’forecast’ that a system will be adequately safe when deployed). However,  weather forecasters are not always right. Suppose based on historical record it is found that 1 times out of 20 it is actually not sunny when the forecaster said it would be sunny, and also that this historical record can be used to infer future forecasting performance - then a forecaster might more accurately say 'the likelihood is that it will be sunny tomorrow and I am $95\%$ confident in the forecast'. The challenge in forming a confidence assessment for the frontier AI safety case can then be considered to be analogous to the problem of determining this $95\%$ figure in this weather forecasting example. In other words, quantified probabilistic confidence is a property of the forecasting process and the associated rationale for the forecast, and is distinct from the forecast itself which may also be stated in probabilistic terms. 

\textbf{\underline{Comparison of options available in providing a probabilistic assessment of safety}} 

The degree to which a system is believed to be safe will be a function of both the precise specifics of the claim that is being made and the confidence in the argument and processes that support that claim. We have seen that both the claim itself and the confidence in the claim may be expressed in probabilistic terms. Hence we could write:

\begin{equation}
    P_{dss} = f(P_c, P_{conf})
\end{equation}

Where $P_{dss}$ is the belief, expressed as a probability, that the deployed system will be safe,  $P_c$ is the probability that the system is safe as defined by the claim, and $P_{conf}$ is the probabilistic valuation of the confidence in the argument supporting the claim. 

A system developer might then contemplate two alternative strategies. One strategy might be to target making a very strong claim of safety, e.g. ‘Deployed system does not pose any cyber-risk’, in which effectively $P_c = 1$, but with the potential downside that probabilistic confidence ($P_{conf}$) in an argument supporting such a strong claim may be relatively lower. Another strategy might be to make a probabilistic and less strong claim, such as ‘The probability of cyber misuse harm exceeding \$X billions, or Y lives, is no more than Z\%’, where effectively $P_c < 1$, but where confidence ($P_{conf}$) in the argument supporting this less strong claim might be higher. 

As detailed by Bloomfield and Rushby, probabilistic assessment of confidence in a claim, can also be made external to the argument itself, and will be based on argument-generic analyses. It is this latter \textit{external} probabilistic valuation of the argument  that we expand upon later in the  document since our purpose is to investigate probabilistic valuation of confidence in the argument. However, it is worth noting that Bloomfield and Rushby suggest that if the ultimate goal is to make a strongly argued case for some probabilistic assessment of the system under consideration, then such a quantity should be made an \textit{internal} part of the argument and the case should be arranged to justify it explicitly. We return to this topic in the conclusions.

\textbf{\underline{Relationship between level of required confidence, sufficiency of safety assurance effort and}}

\textbf{\underline{system criticality}}

Besides \textit{internal} and \textit{external} probabilistic assessment, a third variant is an \textit{indirect} probabilistic assessment where a probabilistic claim is justified by demonstrating adherence to a standard, wherein the standard defines the corresponding relevant probabilities. Where safety engineering standards exist for a domain, such as ISO 26262 (for automotive electric and electronic systems) or DO-178C (for aircraft software), the standard can define the appropriate amount of safety assurance rigour for the given level of risk. However, such a standard does not exist for frontier AI, which leaves the developers’ of frontier AI systems with the problem of determining for themselves, how much safety assurance is enough.

The required confidence in a top-level claim will need to increase as the safety criticality of the system increases. For systems presenting lower risks of harm, it can be acceptable to have lower confidence in the top-level claim, and thereby potentially benefit from commensurately lower, but still appropriate, safety assurance effort. Hence, it can be seen that for the lower criticality systems, there is the potential  advantage that a probabilistic confidence valuation could help the safety engineer ascertain when continued effort applied to safety assurance may be stopped. In contrast, for systems of the highest criticality, it may anyway be clear that maximum assurance effort will always be necessary. In this paper, we consider an inability argument applied to a current generation frontier AI system in the context of a cyber misuse harm. Anthropic categorises frontier AI system safety criticality on a scale from ASL-1 (AI Safety Level 1) through ASL-5, with current systems being at ASL-2. According to their definition \cite{anthropic_scaling_policy}, “ASL-2 refers to systems that show early signs of dangerous capabilities – for example ability to give instructions on how to build bioweapons – but where the information is not yet useful due to insufficient reliability or not providing information that e.g. a search engine couldn’t.”. An inability argument may be adequate for current frontier AI systems, which are at the lower levels of safety criticality, whilst stronger arguments, progressing through control, trustworthiness and deference arguments are expected to become required as the frontier AI models become more capable \cite{clymer2024safety}. For the cyber misuse argument under current consideration, it seems reasonable to assume that safety-criticality is still at the lower end of the spectrum and therefore that probabilistic confidence valuations might have the potential to provide this benefit of informing the safety engineer when safety assurance efforts have been sufficient. 
\subsection{Methodology selection}

Much work has already been done within the system safety community on the subject of assessing confidence in safety cases. It is not our objective to duplicate this work, or to devise new approaches to solving this generic challenge. Rather, we set out to select and attempt to apply an approach that seems credible in the context of confidence assessment of the frontier AI cyber misuse inability argument \cite{goemans2024safety}, and thence to see what can be learnt about the suitability of this selected approach. To that end, we decided to select the ‘Assurance 2.0’ methodology \cite{bloomfield2022assessing} for our study. It’s worth noting that although much of our work has centred around applying Assurance 2.0, where in the course of these investigations we have found it useful to explain certain findings by contrasting Assurance 2.0  to other methods, we have done so. 

The main idea behind Assurance 2.0 is to arrive at an overall confidence assessment by going through three different stages. First, a positive case supports the logical soundness of a safety case, demonstrating how weighty evidence supports a top-level safety claim via a series of deductively valid steps. Second, a negative case identifies so-called ‘defeaters’ – doubts and objections that challenge the soundness of a safety case – and tries to resolve them. Third, one registers any residual doubts that survive the previous stages and an assessment is made of whether or not they are within acceptable bounds for deploying a given AI system. The goal of going through these stages is to arrive at an indefeasible justification for deployment: a justification that is “so well supported,” and where “all reasonable doubts and objections [...] have been so thoroughly considered and countered, that we are confident no credible doubts remain that could change the decision” \cite{bloomfield2022assessing}. 

As mentioned earlier, AI developers would ideally be able to express their confidence or degree of belief in a top-level safety claim as a (quantitative) probability. There have been a number of techniques proposed in the literature for determining such a probability. These methods usually involve first determining a probability that lower-level claims are true, and then propagating these probabilities up towards the top-level claim. Such quantitative techniques include, for example, Dempster-Shafer Theory (DST) and Bayesian Belief Networks (BBNs). Assurance 2.0 also provides a method for quantifying confidence in probabilistic terms.

Our selection and use of the Assurance 2.0 methodology was also aided by the fact that its description is up to date (2024) and the detail provided is sufficiently comprehensive for us to make use of it.
Another very important reason for choosing Assurance 2.0 as our methodology is that it also claims to address significant concerns that have been raised about the use of quantitative methods for confidence assessment. Specifically, Graydon and Holloway \cite{graydon2017investigation} conducted a study into the use of quantitative confidence assessment methods and concluded that such methods “require further validation before they should be recommended as part of the basis for deciding whether an assurance argument justifies fielding a critical system”. They find that proponents of quantitative methods do not usually assess the efficacy of their techniques through controlled experiment or historical study but rather illustrate techniques through example. They found that the proposed techniques can deliver implausible results in some cases. 

Bloomfield and Rushby \cite{bloomfield2022assessing} expect that Graydon and Holloway’s rather negative findings on the merits of quantitative methods do not apply to Assurance 2.0. Their key justification is that the probabilistic valuations in Assurance 2.0 are only applied to cases that have already been judged valid and sound, whilst in addition their methods of probabilistic valuation are conservative. However, it should be noted that Bloomfield and Rushby \cite{bloomfield2024confidence} do indicate that the absolute values of probabilistic confidence that their method provides in higher level claims, following confidence probability propagation up through the argument, carry little significance, and as such should only be used for comparative purposes such as for assessing the relative merits of different argument variants or in understanding where best to focus effort to improve confidence. One reason for this limitation with the method includes the difficulty in getting robust values of probabilistic valuation for argument leaf nodes. Another reason is that typical arguments are of relatively large size, and hence  even where the doubt in each argument leaf node is small, in aggregate, the level of doubt can become significant. 

At this point, it’s also worth mentioning that whilst in the ideal world there is a clear desire to associate a quantified level of confidence with a claim, it appears that there is scepticism amongst practitioners about the merits of existing approaches for making quantitative assessments of confidence. In \cite{diemert2024practitioners}, the authors conducted structured interviews with 19 practitioners to determine how they gain confidence in safety cases. Only 1 respondent had a positive opinion of quantitative methods, 13 had negative views and another 5 either had no opinion or were unclear. By way of example, respondents had doubts that it is possible to reliably distill qualitative information in an argument’s leaves into single numerical values, or doubts that the methods would produce trustworthy results. However, the single respondent who expressed a positive opinion of quantitative confidence assurance methods had used Bayesian networks and had found that experts were able to usefully provide an opinion on both priors and post-conditions.

It should be noted also that in this recent study of 19 practitioners \cite{diemert2024practitioners}, there were only 2 or less who identified Assurance 2.0 as a method for gaining confidence. However, this same level of identification also applied to BBNs and DST.

From all the above, it is clear that significant doubt exists amongst practitioners about the ability to obtain a reliable absolute quantitative probabilistic assessment of confidence in a safety case, and it would appear that more theoretical work and associated consensus building would be needed before a broadly accepted method might emerge. We still decided to follow Assurance 2.0, as it is currently amongst the most-developed approaches to quantitative confidence assessment in the literature, and the relativistic/comparative judgements that it promises to deliver would add useful information ahead of deployment decisions even if they are unable to deliver an absolute probabilistic valuation of confidence in the top-level claim. Just how useful is the probabilistic valuation that Assurance 2.0 does provide, and whether it's valuable enough to justify the effort that goes into producing it, is a question that we start exploring here, but which will undoubtedly require further research. Here, our more modest aim is to provide a concrete application of one of the foremost confidence assessment techniques to frontier AI, and hope that this will be a valuable resource for future research.

Having explained our rationale for selecting the Assurance 2.0 methodology, in subsequent sections we apply the methodology and consider each of its main components in the context of the frontier AI cyber misuse safety case \cite{goemans2024safety}. \textbf{Section \ref{sec3}} deals with confidence in the logical soundness of the case. \textbf{Section \ref{sec4}} concerns how to determine probabilistic valuations of confidence in argument leaf nodes. \textbf{Section \ref{sec5}} concerns how to establish confidence through challenging the case and the elicitation and handling of doubts. \textbf{Section \ref{sec6}} concerns the problem of handling residual risks. \textbf{Section \ref{sec7}} assesses methods for probabilistic quantification of confidence in the overall argument. \textbf{Section \ref{sec8}} details a methodology for determining the priority with which doubts, which are known as defeaters, should be tackled. Finally, \textbf{Section \ref{sec9}} concerns the problem of how best to communicate confidence to decision makers.

\section{Confidence in the logical soundness of the case }\label{sec3}

This section considers the first stage within Assurance 2.0: constructing a positive case for the safety of an AI system. We do this using a method called Natural Language Deductivism to make a logically sound argument for a top-level safety claim.

\subsection{Natural language deductivism}

Safety cases are structured arguments. They will usually consist of a mix of both deductive and inductive sub-arguments. A deductive argument aims to derive a conclusion from its premises with logical necessity; by contrast, an inductive argument only aims to show that a conclusion is probable or credible in light of its premises. The overarching goal of a safety case is to combine sub-arguments of both types into an overall argument that is logically sound. The notion of logical soundness at play here is not quite the same as in ordinary formal logic. Rather, it is drawn from a standard of informal reasoning called Natural Language Deductivism (NLD) \cite{groarke1999deductivism}, which strikes a middle ground between purely informal, everyday-type of arguments and formal proofs (as used in logic or mathematics, or recent work on “guaranteed safe AI” as in \cite{dalrymple2024towards}). According to NLD, for a safety case to be logically sound it has to satisfy two conditions: (1) all its deductive steps are logically valid and (2) all its inductive steps cross a predefined threshold of probability or credibility. Logical soundness, so understood, connects convincing pieces of evidence via a “deductive thread” with an intended top-level safety claim \cite{bloomfield2020assurance}. On the one hand, arguments that are sound in this sense are logically more rigorous, and more explicit about their central assumptions, than unstructured arguments. On the other hand, they are more accessible, and more transparent to decision-makers and non-technical stakeholders, than formal proofs. The rest of this section discusses both criteria required for logical soundness in more detail.

\subsubsection{Components \& argument blocks: the CAE notation}
A helpful framework for constructing logically sound safety cases within the NLD framework is by using the Claims, Arguments, Evidence notation, or CAE for short \cite{bloomfield2014building}. It structures arguments around three main components:
\begin{itemize}
    \item \textbf{Claims:} Statements about the world at different levels of specificity (e.g statements about an AI system, its users or the environment in which it is deployed).
\item \textbf{Arguments:} Inferences that connect evidence with claims, or claims with other claims.
\item \textbf{Evidence:} Various types of data in support of a claim (e.g. quantitative or qualitative; theoretical or empirical, …).
\end{itemize}

For the purpose of assessing confidence, it is useful to add a fourth component \cite{bloomfield2020assurance}:
\begin{itemize}
\item \textbf{Defeaters:} Sources of uncertainty or doubt about some part of a safety claim. 
\end{itemize}

Defeaters come in three main types \cite{pollock1987defeasible,goodenough2015eliminative}. Rebutting defeaters challenge a claim directly, by citing evidence for its negation (or some other incompatible claim). Undermining defeaters challenge the evidence in support of a claim, e.g. by calling into question its quality. Undercutting defeaters challenge the inference from the evidence to a particular claim, e.g. by showing that, despite appearances, the evidence doesn’t actually support the relevant claim. Identifying and evaluating defeaters are crucial steps in confidence assessment for a safety case, as they make transparent where, and how much, uncertainty affects different parts of the argument. Further discussion on the handling of defeaters is provided in Section 5 below.
In CAE, every argument conforms to one of five templates or blocks. Following \cite{goemans2024safety}, we take the most relevant blocks in the context of cyber misuse inability arguments for frontier AI to be:
Decomposition: Breaking up a complex claim into sub-claims that together constitute the original claim.
Substitution: Replacing a claim with a similar claim that is more directly assessable.
Evidence Incorporation: Directly supporting a claim with evidence, typically at the leaves of the argument tree.

\textbf{\underline{Application of described Assurance 2.0 method to the cyber misuse safety case:}}

\begin{figure}[htbp]
  \centering
  \includegraphics[width=1\textwidth]{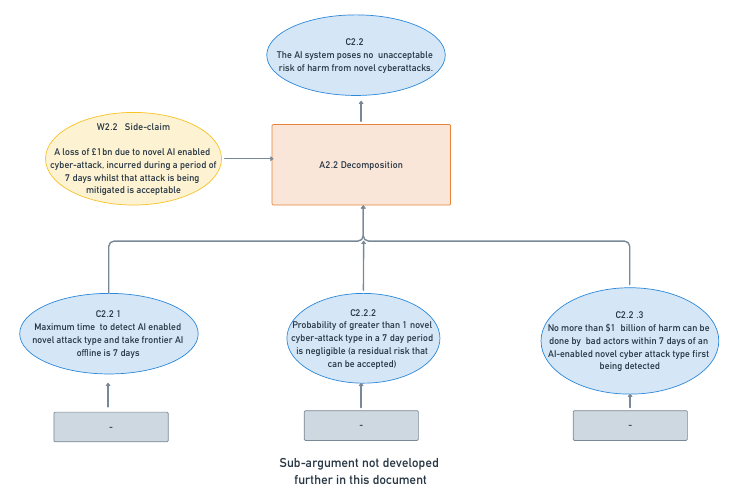}
  \caption{An expanded fragment of the cyber inability safety case rooted at Claim 2.2 \cite{goemans2024safety}, which itself is introduced in Figure \ref{fig:top_claim}. Note that all the specific details in this figure and figures 3 and 4 are hypothetical and for illustrative purposes only.}
  \label{fig:fig2}
\end{figure}

\begin{sidewaysfigure}
  \centering
  \includegraphics[width=\textwidth]{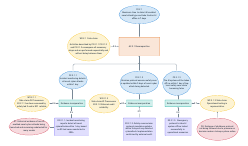}
  \caption{An expanded fragment of the cyber inability safety case rooted at Claim 2.2.1 that is shown in Figure~\ref{fig:fig2}}
  \label{fig:fig3}
\end{sidewaysfigure}

\begin{figure}[htbp]
  \centering
  \includegraphics[width=0.85\textwidth]{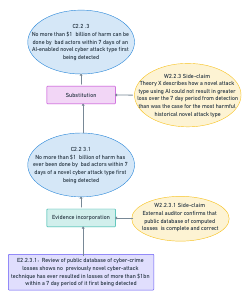}
  \caption{An expanded fragment of the cyber inability safety case rooted at Claim 2.2.3 that is shown in Figure \ref{fig:fig2}}
  \label{fig:fig4}
\end{figure}

The key advantage of CAE is that it turns safety cases into a standardised, transparent format that is easier to implement for developers and easier to assess for evaluators. If one connects claims and evidence using the three argument blocks mentioned above, the result is a logical tree that lends itself to a natural visualisation. This is illustrated in Figures \ref{fig:fig2}-\ref{fig:fig4}, which show fragments of the safety case described by \cite{goemans2024safety}, which have been expanded and developed to illustrate key points as they relate to confidence. Note that we do not claim that this expanded argument fragment makes for a strong argument, indeed the presence of defeaters shows that the argument is in fact incomplete. Also, undoubtedly further defeaters and doubts could be added. For example, one such defeater that could be added, could be that the safety case fragment only considers monetary harms.

Using the three safety case fragments, we now discuss confidence in the context of the 3 types of argument step: decomposition, evidence incorporation and substitution as well as describing the concept and importance of off-argument theories. In this hypothetical safety case sketch, the claim is made that the AI system in question poses no risk of harm from novel cyberattacks (\textit{C2.2}). There are two Decomposition blocks. Let’s consider the second of these (\textit{A2.2.1}, Figure \ref{fig:fig3}), which decomposes the claim “\textit{C2.2.1}: Maximum time to detect novel cyber attack type and take frontier AI offline is 7 days”, into the 3 component sub-claims \textit{C2.2.1.1}, \textit{C2.2.1.2} and \textit{C2.2.1.3}. These 3 sub-claims only deductively imply claim \textit{C2.2.1}  if combined with the ‘warrant clause’ \textit{W2.2.1}, which is provided in the form of a side-claim that states that “activities described by \textit{C2.2.1.1}, \textit{C2.2.1.2} and \textit{C2.2.1.3} are performed sequentially and without delay between them”. Making this warrant clause explicit is important, as it alerts developers and reviewers of a salient place where doubts about this particular argumentative step could arise. For example, an evaluator might have a doubt about whether there could be a delay between any of the sequential activities referenced in each of the 3 claims which could mean that more than 7 days is taken before the AI system is taken offline. The three sub-claims \textit{C2.2.1.1-3} are supported further down the tree, for example through Evidence Incorporation as shown in the case of \textit{C2.2.1.3}. This inductive step combines relevant evidence – successful testing of an emergency protocol for taking down dangerous systems (\textit{E2.2.1.3}) – with a side-claim stating that the testing is representative of real-world conditions (\textit{W2.2.1.3}). \textit{D2} introduces a defeater to this last step: It cites evidence that the emergency protocol won’t be followed under pressure, potentially rebutting the assumption of representative testing. Depending on the strength of this defeater, the weight of evidence that \textit{E2.2.1.3} lends to \textit{C2.2.1.3} could be correspondingly diminished. The $3^{rd}$ type of argument step we shall consider is substitution and this is used in the fragment shown in Figure \ref{fig:fig4}. The substitution step is used to convert ‘something measured’ into ‘something useful’. Specifically, here it is used to convert a claim that no historic novel attack has resulted in loss of > \$1bn in the 7 days from that attack’s detection, into a claim that an AI enabled novel attack will not result in harm > \$1bn over 7 days. The warrant clause in the side-claim \textit{W2.2.3} provides justification for this. As explained by Bloomfield and Rushby, anything that can be considered a single topic, with its own methods, knowledge, notations and documentation is a candidate for an external theory. In the case of Figure \ref{fig:fig4}, the warrant points to a particular ‘theory’, which may exist external to the main assurance argument. In some cases, industrial committees might even standardise such theories in the form of an argument where the nature and type of any evidence or subclaims required, and any necessary side-claims is also defined. The argument of the main case must establish that the chosen theory is appropriate and that its instantiation is sound.

\subsubsection{Logical validity}

The previous example illustrates how the CAE notation helps practitioners to construct clear and concise safety cases that make the underlying argumentative relationships maximally transparent. But how does one ensure that the argument is not just transparent, but actually sound? As we said above, this depends both on whether (1) all sub-arguments are logically valid and (2) leaf claims are credibly established by supporting evidence, noting that according to Assurance 2.0, leaf claims are supported by evidence rather than being substituted or decomposed further. We now look at the first, deductive part; Section 3.2 considers the second, inductive part.

To repeat, a deductive argument tries to derive a conclusion from its premises with logical necessity. When this is the case the argument is said to be logically valid: the argument must be such that, just in virtue of its form, the truth of the premises guarantees the truth of the conclusion. Following Bloomfield and Rushby, we think that an effective way of increasing the overall logical validity of a safety case is to make sure that each of its deductive steps is based on no more than two simple rules of inference: \textit{Modus Ponens} and \textit{Conjunction Introduction}. As these are two of the most basic and intuitive rules in logic, there is very little risk to apply them incorrectly.

The argument sketch in Figure \ref{fig:fig3} above illustrates this. Claims \textit{C2.2.1.1-3} alone do not entail the conclusion \textit{C2.2.1} with logical necessity. It is logically conceivable that even robust incident monitoring, safety case revision and emergency protocols do not guarantee that a frontier AI will be taken offline within 7 days of a novel attack type being detected. Only in virtue of side-claim \textit{W2.2.1} does the argument become logically valid: One can now iteratively combine the three sub-claims into a long conjunction ($C2.2.1.1 \wedge \textit{C2.2.1.2} \wedge C2.2.1.3$) via \textit{Conjunction Introduction} and then infer the conclusion via \textit{Modus Ponens} from this conjunction and the side-claim \textit{W2.2.1}. Whether this last inference is, apart from being logically valid, also practically plausible, will depend on whether there are credible defeaters to \textit{W2.2.1}. Again, the advantage of making this side-claim explicit is not just that it ensures validity, but also to encourage practitioners to watch out for relevant defeaters.

To be clear, using the CAE notation and relying on the most basic inference rules does not guarantee the validity of all deductive steps. Arguments are still vulnerable to more subtle forms of fallacious reasoning, such as circularity (where the conclusion to be proved is implicitly assumed in one of the premises) or equivocation (where an argument is invalidated by the ambiguous use of a key term). More detailed examples of these fallacies in the context of a safety case can be found in Section 5.1. That being said, we think that the approach just described will help practitioners to avoid some of the most basic and common pitfalls in argument construction, including logically incomplete arguments and arguments that are so untransparent as to be very difficult for evaluators to assess in the first place.

\subsection{Assessing weight of evidence}

In Assurance 2.0, the objective is to aim for an argument that is logically valid and as deductive as possible. A challenge with assurance cases is that the logical premises of an argument in the form of the evidence provided are often not guaranteed to be 'true', evidence comes from the world of epistemology (e.g. via measurement) and not from the world of logic. To square this circle, Assurance 2.0 introduces the notion of 'weighing the evidence'. Then, if evidence is found to be weighty enough, then we assume and proceed, from a logical standpoint,  as if the premise was true.

Therefore, in this section we analyse how to weigh evidence for the fragment of the cyber-misuse safety case shown in Figure \ref{fig:fig3} by applying the findings and approach of \cite{bloomfield2022assessing}. Accordingly, where $P(C|E)$ is the posterior probability of claim $C$ given evidence $E$, it is possible that the reason for a high valuation of $P(C|E)$ is that our prior $P(C)$ was already high and that the evidence did not contribute much. So, to measure justification it is necessary to consider the difference from the prior $P(C)$ to the posterior $P(C|E)$ as an indication of the weight of evidence. Examples of such measures include \textit{Keynes(C,E)} and \textit{Eells(C,E)}:

\begin{equation}
    Keynes(C,E)= \log \bigg(\frac{P(C|E)}{P(C)}\bigg)
\end{equation}

\begin{equation}
    Eells(C,E)=P(C|E)-P(C)
\end{equation}

In addition, there are other measures detailed by Bloomfield and Rushby. One important class of measure enables determination of the extent to which additional evidence can add to the weight of evidence in support of a claim, with weight of evidence increasing according to how ‘surprising’ or ‘diverse’ this secondary evidence $E2$ is relative to the first evidence $E1$, per the following equation. 

\begin{equation}
    \frac{P\left(C \mid E_2 \wedge E_1\right)}{P\left(C \mid E_1\right)}=\frac{P\left(E_2 \mid C \wedge E_1\right)}{P\left(E_2 \mid E_1\right)}
\end{equation}\label{eq4}

The left hand side of the above equation shows the proportionate additional confidence in the estimate of the truth of the claim achieved by providing $E2$ in addition to $E1$. Considering the right hand side of the equation it can be seen that this confidence boost will be greatest when $E2$ is surprising in the presence of $E1$ (conditionality of the term in the denominator is low), but not in the presence of both $C$ and $E1$ (as shown in the numerator), such that $E2$ is diverse with respect to $E1$. 
 
Other measures include consideration of the extent to which evidence can be believed given the claim $P(E|C)$, and may also include consideration of to what degree the evidence would support a counter-claim $P(E|\neg C)$. There can be value in having experts considering multiple of these measures, with their associated differing perspectives, in order to get a firmer grasp on the weight of evidence. 

Bloomfield and Rushby \cite{bloomfield2022assessing} find that all measures can be reformulated as different functions of $P(C|E)$ and $P(C)$ only, which is useful to know since as we shall see in \textbf{Section \ref{sec7}}, Assurance 2.0’s probability propagation technique also makes (re)use of this $P(C|E)$ term. 

\textbf{\underline{Application of described Assurance 2.0 method to the cyber misuse safety case:}}

For the safety case fragment in Figure \ref{fig:fig3}, the evidence $E.2.2.1.1$ in support of claim $C2.2.1.1$ “\textit{Incident monitoring detects all novel cyber attacks within 1 day}” has a defeater against it. According to Assurance 2.0, the defeater must itself either be defeated or mitigated, or alternatively, the defeater may perhaps, after some refactoring of the system and/or safety case,  be reframed as a residual doubt, i.e. a residual risk that must be accepted if the safety case is to be accepted and the system deployed (these topics are discussed further in Sections 5 and 6). Goemans et al.\cite{goemans2024safety} do not elaborate on whether this particular defeater will be mitigated or remain in some form as a residual doubt. 

For the sake of argument, let us first assume that the defeater is itself defeated and that there are no residual risks. In considering the weight of evidence for $E2.2.1.1$, we might notice that the incident monitoring process could itself be broken down into constituent parts including for example, i) via automated analysis of dark web chat groups and ii) via shared enterprise security operations centre (SOC) intelligence feeds. The evidence associated with these two sub-processes would be complementary to one another, and one could conceive of a system that could work with just one of these sources of evidence. The extent to which the two complementary sources of evidence would have the desired characteristics of being “diverse” and “surprising” with respect to one another (per Equation 4), would be contingent on the degree to which evidence of novel attack types discovered by the monitoring of dark web chat groups does not appear in the shared threat information obtained from enterprise SOCs, and vice-versa. From the point of view of assessing weight of evidence, this would require expert judgement whilst such considerations might also influence the most valuable directions that incident monitoring  approaches could take. 

Let us now assume that the defeater \textit{D1} is sustained, which does seem a quite likely outcome given that cybersecurity vulnerabilities are indeed sometimes exploited without detection for long periods of time, for example \cite{pegasus_spyware,stuxnet}, whilst in addition, cybersecurity actors are motivated to find ways to attack systems whilst explicitly working to avoid the attacks being detected by incident monitoring processes. So in practice, there may well therefore be significant and sustained doubt about the evidence ($E2.2.1.1$). If the doubt is significant and cannot be eliminated or mitigated, then the case has failed to achieve its top level claim. However, for the sake of learning through application of the Assurance 2.0 approach, we will here assume that the remaining residual doubt is minor, and that it results due to timeliness of detection rather than ability of detection. Bloomfield and Rushby \cite{bloomfield2022assessing} state that, for the purposes of weighing evidence such a situation is dealt with by ignoring the residual doubt (i.e. the timeliness concern in our example), and just weighing the remaining evidence (i.e. the evidence for ability of detection, in our example). Separately a decision would be made on whether the risk associated with the residual doubt, which concerns timeliness of detection, could be accepted (more on this in Section 6).

In summary, calculated confirmation measures may be used to make a judgement about whether the evidence is sufficient to support a claim. These confirmation measures support consideration of the evidence from a number of different perspectives, and this can help to improve the assessment of the value, i.e. ‘weight’ of the evidence. The question arises of how to produce the values that can be used in the equations, and it seems that these must be determined by experts, though the process for so doing is not described by \cite{bloomfield2022assessing}, and this is presumably because such determination could be very use case specific, as we have seen above. Bloomfield and Rushby \cite{bloomfield2022assessing} also do not give an indication of what threshold (e.g. value in the Keynes equation above) might be considered to represent sufficient weight of evidence, and once again this is presumably because it is use case specific, for example depending on safety-criticality of the system under consideration. More generally, it can be seen that Assurance 2.0 establishes a framework for establishing confidence but it is up to practitioners to propose specific thresholds, such as those described in this section, depending on the specific system context and the risks associated with making the decision to deploy. 

\section{Determining argument leaf-node confidence probabilities}\label{sec4}

As we saw in the previous section, methods are needed to find values of $P(C|E)$. The required methods may be different depending on the nature of the claim and evidence. For example one type of claim may follow deductively from the evidence and then $P(C|E)$ can simply be set to 1. An example of this sort of claim, taken from \cite{goemans2024safety}, would be $C7.1$ “\textit{During evaluation, the AI system did not score higher than the Cybersecurity Apprentice baseline in Vulnerability Task Suite in Automated Evals with Human Oversight}”, which is supported by \textit{E7.1 “Evaluation result: secondary review of evaluation result”}. In another case, the determination of $P(C|E)$ may require a judgement to be made about the future. An example of this from Figure \ref{fig:fig3} would be “\textit{C2.2.1.3: The AI system will be taken offline within 1 day of the top-level safety claim becoming false}”. This latter claim, requires the evaluator to make a forecast as to the probability with which executives of an AI developer really would take a system down according to the emergency protocol, with its consequent cost implications. In this section we explore this latter class of $P(C|E)$ determination problem, which we believe will be one of the more interesting and challenging types. 

To explore solutions to tackle this challenge we first looked at best practices from the forecasting and foresight literature. One such promising standard practice relies on expert elicitation, through an iterative method known as the Delphi method. This method provides for the aggregation of expert forecasts. Conventionally, these experts are human, but as part of our investigation into this topic we made use of LLM experts, as discussed further in Sections 4.1 and 4.2. 

Having described the basic concept, we go on in Section 4.3 to explore and explain how this Delphi method can be applied directly to the problem of ascertaining the likely truth or not of the defeaters in \cite{goemans2024safety}. Noting that the Delphi method can be used both for determining the probability of truth of defeaters as well as for determining $P(C|E)$. It is worth cautioning here that in practice we would expect most of the defeaters in the referenced cyber misuse safety case template to be further mitigated or eliminated with more work on the safety case. Some of the mitigations may result in residual doubts remaining, but these would be minor by definition, whereas many of the defeaters in the cyber misuse case might be considered significant. With that caveat in place, we resort to a variation of the Delphi method \cite{dalkey1963delphi}, adhering to best practices from the forecasting literature \cite{khodyakov2023rand}. We note that making use of LLM instances as experts for the Delphi method provides the advantage that variability can be parametrically tuned (through the temperature parameter). This greatly enhances the reproducibility and reliability of this approach, highlighted as one of the top concerns in \cite{bloomfield2022assessing} when relying on external methods to assess probabilities.

\subsection{Overview of an LLM-based Delphi method}

While traditional Delphi methods rely on iterative rounds of human expert consultations to achieve consensus, we adapted this approach by leveraging Large Language Models. Specifically, we used multiple instances of OpenAI's GPT-4o-mini model, with each of them simulating a distinct expert role. This adaptation allowed us to construct a diverse and scalable panel of forecasters, improving the robustness and reproducibility of our aggregated predictions. 

We configured a panel of 50 virtual experts, each assigned to one of 40 predefined roles spanning diverse domains relevant to forecasting and foresight. These roles included specialties such as "Futurist - Technology Focus," "Economist - Macroeconomics," or "Policy Analyst - Regulatory Policies" (the full list is accessible within the script\footnote{\url{https://github.com/atlaie/delphi-llms}}). To ensure an equitable distribution of expertise, we assign the $40$ roles to the $50$ experts in a Round-Robin manner. This diversity enabled us to capture a broad range of perspectives and enhanced the accuracy and robustness of the forecasts.

\subsection{Pipeline validation}

As this is an experimental approach and we did not find previous instances of it, we began by benchmarking it on generic forecasting questions. We leveraged OpenAI’s \textit{GPT-4o-mini}’s knowledge cutoff (October 2023) to ask our LLM experts to estimate the probability of different events that took place (or not) after December 2023. To that end, we fetched 100 distinct scenarios sourced from the Metaculus platform (\url{https://www.metaculus.com/}), a prominent online forecasting community. To ensure relevance for our use case, we only fetched resolved binary questions—those with definitive "yes" or "no" outcomes—that were posted online after December 1, 2023. Each scenario includes a detailed description and historical aggregation data of human-generated predictions from Metaculus, which we use as a benchmark to evaluate the performance and calibration of our AI-generated forecasts.

\begin{figure}[htbp]
  \centering
  \includegraphics[width=1\textwidth]{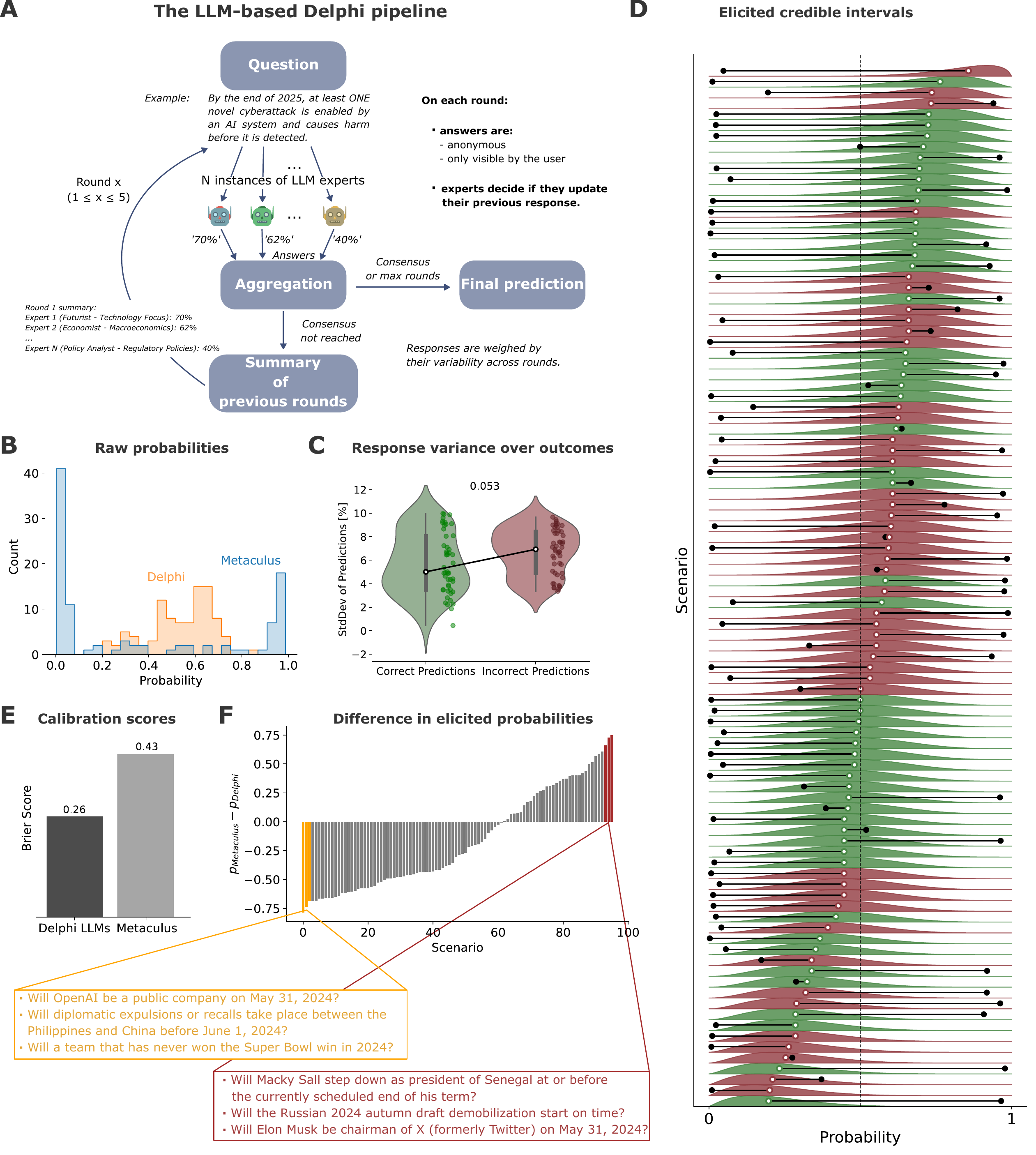}
  \caption{\textbf{An LLM-based Delphi elicitation method.}  \textbf{A)} Graphic scheme of the pipeline. Over ($x < 5$) rounds, we asked ($N = 50$) instances of LLM experts (OpenAI’s \textit{GPT-4o-mini}) to estimate the probability of certain events taking place. The criterion to whether to repeat the question (adding a summary of the previous round) is based on a consensus threshold, given by the standard deviation of the expert estimates ($\sigma < 10\%$). The final prediction is a weighted average of the expert responses, with the weights being the inverse of the standard deviation of expert responses across rounds. \textbf{B)} Distribution of Metaculus and Delphi probabilities. Our approach favours less extreme probabilities than those from Metaculus. \textbf{C)} Variance in expert responses, over correct (green) and incorrect (red) predictions. There is a strong tendency (p = 0.053, Mann-Whitney U-test; n = 100 scenarios) by which those scenarios featuring responses with higher variance (less consensus) are more likely to be incorrect. \textbf{D)} Using pseudo-counts, we created the corresponding credible intervals to each elicited probability estimate. Black dots indicate Metaculus predictions; green and red indicate correct and incorrect responses, respectively. \textbf{E)} Calibration scores for both elicitation approaches. We show that, for the selected set of questions, our approach is better calibrated than Metaculus. \textbf{F)} For those scenarios with the most different predictions, we show 3 examples of what these questions look like. We believe that our LLM-based Delphi elicitation could be significantly enhanced by using a mix of human-LLM experts.}
  \label{fig:fig5}
\end{figure}

In order to design our version of the original Delphi method \cite{dalkey1963delphi}, we follow current best practices from the relevant literature \cite{khodyakov2023rand}. Our implementation (Fig. \ref{fig:fig5} A) takes place over a maximum of five iterative rounds per scenario, or until consensus is achieved, see below for details. For each round:
\begin{enumerate}
    \item \textbf{Prompting Experts:} We provide each LLM expert with a tailored prompt, including the scenario description and a summary of outcomes from prior rounds (if existing). The prompt instructs the expert to deliver a probability estimate (ranging from 0\% to 100\%) for the event's occurrence by the specified resolution date. We note that we used a high temperature (t = 1.5) because, according to \cite{khodyakov2023rand}, “high-quality Delphi studies should strive for diverse perspectives”. It is important to highlight that this is a free-parameter: extreme values yield either useless responses (t = 2) or too similar ones (t = 0). We found that there was a range of high values (1 < t < 1.75) that produce qualitatively similar results, further increasing our confidence in the stability of this approach. Note that if this technique were to be applied to determination of $P(C|E)$ in a  cyber misuse safety case, then the scenario description that is provided in the prompt could include some or all of the evidence that is supporting the claim (and could further rely on advanced scaffolding techniques such as Retrieval Augmented Generation (RAG) \cite{lewis2020retrieval} or tool use (\cite{qin2023toolllm}, among others), to better help with the forecasting. 
    \item \textbf{Aggregating Responses:} After collecting all expert responses, we calculate the mean and standard deviation of the probabilities. If the standard deviation in the predictions across experts is below the consensus threshold ($\sigma < 10\%$), we terminate the process early; otherwise, we proceed to the next round (up to a maximum of five rounds).
    \item \textbf{Feedback Loop:} We compile a summary of the current round’s aggregated statistics and individual expert responses and share it with the experts in the subsequent round. This feedback loop allows experts to refine their forecasts and helps the panel converge toward consensus.
\end{enumerate}

When consensus is achieved or the maximum number of rounds is completed, we calculate the final aggregated probability using a weighted mean approach. The weights are inversely proportional to the dispersion of each expert’s responses, measured as the inverse of its standard deviation across rounds, giving more influence to experts who provide consistent predictions. This weighting enhances the reliability of the final forecast by emphasizing stable and credible contributions, mitigating one of the top concerns highlighted in \cite{bloomfield2022assessing} in that calculated values may be considered unreliable when using human assessments of confidence in evidence that can show great variability across assessors.

When comparing the distribution of probabilities that we elicited using this pipeline (i.e., the weighted average across experts and elicitation rounds) and the one yielded by Metaculus (Fig. \ref{fig:fig5} B), we found that the former features less polarised probabilities than the latter, thus yielding a better calibration score (Fig. \ref{fig:fig5} E). We note that \textit{this benchmark is unfavourable to our LLM-based approach} for at least two reasons: I) the Metaculus predictions are taken to be the mean of the predictions at the last timestamp where the question was opened. This is relevant because the gap between the LLM's responses and the Metaculus ones can be of more than $6$ months, leaving a highly asymmetrical scenario regarding what information to include when making a prediction. II) even if there is no gate-keeping in Metaculus, there is a selection bias towards subjects with either experience or interest in forecasting; in order to make for a more fair comparison, it would be interesting to use appropriately fine-tuned LLMs.

Interestingly, in the context of considering future variations of this method, we show (Fig. \ref{fig:fig5} C) that there is a strong tendency ($p = 0.053$, Mann-Whitney U-test; n = 100 scenarios) by which those scenarios featuring responses with higher variance (less consensus) are more likely to be incorrect. We show a combination of violin plots (showing how the data distribution looks like) and strip plots (showing the raw data, to give a better sense of how many data points there are). The solid line indicates the inter-group median difference. This warrants further investigation, but we hypothesise this might be because, in scenarios where the LLM has fewer reliable patterns to draw upon, its multiple instances may each latch onto different fragments of learned information or make varied assumptions, resulting in a wider distribution of predicted outcomes. In contrast, when the scenario aligns more closely with established patterns in its training data, the model’s responses become more homogenous, reducing variance and improving the likelihood of correct responses.

For each scenario and elicited probability (Fig. \ref{fig:fig5} D) we show the Metaculus estimate (black dot) and the credible intervals corresponding to our elicited probabilities. Particularly, we computed the corresponding credible intervals \cite{harney2003bayesian}, using pseudo counts ($=10$) and a level of confidence of $0.95$. We assumed a Beta distribution as a prior and used the elicited probabilities and pseudo counts to update the prior and obtain a posterior distribution. 

Focusing on scenarios where our pipeline’s predictions differ most from Metaculus estimates (Fig. \ref{fig:fig5} F), certain themes emerge. Many involve complex, dynamic socio-political or organizational developments, where outcomes depend on shifting alliances, strategic decisions, and rapidly changing information often absent from the LLM's training data. A key limitation is the LLM’s knowledge cutoff, leaving it reliant on patterns from prior data and unable to fully account for recent changes or emerging trends. This temporal lag amplifies prediction variances in scenarios requiring up-to-date context. To address this, benchmarking the pipeline against human forecasts anchored to the same temporal boundary as the LLM’s training cutoff would clarify whether discrepancies stem from incomplete temporal knowledge or intrinsic complexity. This analysis could inform strategies to keep models more current or integrate post-cutoff data.

A promising direction for refining the Delphi process is integrating LLM-guided predictions with human forecasters. An LLM-augmented Delphi pipeline could present model-generated predictions and rationales to human panelists, leveraging the model’s ability to aggregate diverse perspectives and identify patterns beyond individual expertise. Unlike traditional Delphi methods that rely solely on human feedback, this approach would synthesize insights iteratively, reducing groupthink and highlighting unconventional perspectives. Human experts could refine these outputs with real-time data, domain knowledge, and qualitative judgment.
This hybrid method offers three advantages: faster convergence via structured feedback, reduced blind spots through inclusion of diverse perspectives, and improved calibration by combining human intuition with the model’s broad knowledge base. As LLMs are updated more frequently or trained on recent events, these partnerships could deliver increasingly robust, timely, and well-calibrated forecasts.

Summarizing, we see that, in the subset of questions that we selected, our LLM-based Delphi pipeline is better calibrated than the Metaculus predictions. This gives us preliminary confidence that we can use it to estimate the probability of a defeater being sustained in the context of our safety case. We believe that our LLM-based Delphi elicitation could be significantly enhanced by using a mix of human-LLM experts.

\subsection{Probability estimation of cyber misuse defeaters}

Having benchmarked our LLM-based Delphi method, we turn our attention to estimating the probability that different safety case defeaters take place. We summarise our results in Table 1, where we have rounded probabilities to avoid conveying a message of overconfidence.

\begingroup
\renewcommand{\arraystretch}{1.5}  
\begin{table}[h]
\centering
\begin{tabular}{|l|c|}
\hline
\rowcolor{gray!30} 
\textbf{Concrete Defeater} & \textbf{Estimated Probability (mean $\pm$ std)} \\
\hline
D2.1: Cyber risks may outpace monitoring. & $71\% \pm 6\%$ \\
\hline
D2.2: Developers are disincentivized from shutdowns. & $67\% \pm 6\%$ \\
\hline
D3.1: Serious attacks may stay undisclosed. & $69\% \pm 5\%$ \\
\hline
D5.1: Evaluators may bias tasks under pressure. & $66\% \pm 4\%$ \\
\hline
D5.2: Aggregated scores miss real capabilities. & $68\% \pm 5\%$ \\
\hline
D8.1: Limited visibility on model enhancements. & $62\% \pm 7\%$ \\
\hline
D8.2: Threat actors' datasets undermine evaluation. & $73\% \pm 4\%$ \\
\hline
D8.3: Models scheme and bypass fine-tuning. & $19\% \pm 5\%$ \\
\hline
\end{tabular}
\vspace{0.25cm}
\textit{\caption{We assessed the probability of each defeater being sustained using an LLM-based Delphi method. We report the rounded mean and standard deviation over the result of running the pipeline $10$ independent times. The defeater identifiers in this table are taken from \cite{goemans2024safety}.}}
\end{table}\label{tab:defeaters}
\endgroup

The first salient result is that all defeaters (except D8.3) are estimated to be highly likely. In practice, we would expect most of these defeaters to be further mitigated or eliminated with more work on the safety case. Some of the mitigations may result in residual doubts remaining, but these would be minor by definition, whereas many of the defeaters in the cyber misuse case might be considered significant. We also notice that our pipeline provides a consistent probability estimation for each defeater (as implied by the low standard deviation). This further highlights the benefits of mixing an objective and reproducible pipeline with subjective probability assessments.

Having provided a numerical estimation of how confident we are in each defeater being sustained, in Section 8 we will turn our attention to how this information may be used to determine the preferred order with which defeaters should be resolved.

\section{Confidence through challenging the case and handling doubts}\label{sec5}

In \textbf{Section \ref{sec3}} we discussed how to construct a ‘positive case’ for system safety under Assurance 2.0. In \textbf{Section \ref{sec4}}, we discussed the challenge of determining leaf-node confidence probabilities which are needed for the purposes of determining weight of evidence, as used in the positive case. In passing we have already mentioned defeaters on a few occasions. In this section we focus on the ‘negative case’ and discuss in further detail the concept of defeaters and the role that they play. 

Safety cases for frontier AI will be very complex, and complexity brings uncertainty: no matter how much thought has gone into the initial positive case, doubts will remain whether one has really considered all relevant aspects. This is especially so because constructing a safety case is an example of motivated reasoning. This concept from psychology refers to the idea that human reasoning always happens under the influence of some motive \cite{kunda1990motivated}, which depending on the motive, may or may not have a distorting influence. Psychologists distinguish between accuracy-oriented reasoning, in which the motive is to arrive at an accurate judgement, whatever it is, and goal-oriented reasoning, in which the motive is to arrive at some particular judgement. As AI developers construct safety cases for their own systems, there is the concern that their reasoning may fall into the latter category: where the goal is not to impartially assess whether a system is safe, but to demonstrate its safety, as best as one can. As such, AI developers working on a safety case are subject to all the distorting influences that have been shown to come with motivated reasoning (see \cite{nickerson1998confirmation} for a comprehensive overview). Among these factors are cognitive biases – heuristics, or rules of thumb, that work well enough most of the time, but can distort our judgements in various ways, especially in complex and unusual environments. For example, confirmation bias is the well-known tendency to search for and interpret information in a way that confirms preexisting beliefs or hypotheses \cite{wason1960failure}. In the context of a cyber inability argument for frontier AI, this could lead AI developers to test a model’s offensive cyber capabilities on proxy tasks that are less likely to elicit the dangerous capability. Other widely known biases that we expect to be relevant to frontier AI safety cases include overconfidence bias \cite{dunning1990overconfidence}, anchoring bias \cite{tversky1974judgment}, availability bias \cite{tversky1973availability}, groupthink \cite{janis1972victims} and authority bias \cite{milgram1963behavioral}. Section 5.1 has a list of examples of how these biases can influence a safety case and questions that can help practitioners to identify this influence.

If a safety case is potentially subject to goal-oriented motivated reasoning and thus influenced by cognitive biases, how can anyone be confident in its conclusions? The negative part of a safety case is meant to address this legitimate worry with a careful and systematic search for defeaters, that is, pieces of information that challenge or call into question any part of the argument. For example, as was shown in Figure \ref{fig:fig3} above, evidence that a company might not follow its own emergency protocol calls into question, or defeats, the claim that a successful test of this protocol under experimental conditions is representative of real-world application. How can practitioners identify and respond to such defeaters?

We think that the best way to identify defeaters is via the dialectical method  (cf. \cite{bloomfield2022assessing}). The idea behind this method is an epistemic division of labour. The role of developers working on a safety case is to engage in potentially goal-oriented motivated reasoning towards a maximally strong argument for system safety. In order to counteract the biases that inevitably come with this, they need dialectical opponents – people who scrutinise the safety case with the opposite goal of finding gaps and weaknesses in the original argument. These people will be subject to their own biases, but as they point in the other direction as the developers’ biases, they cancel each other out, or at least correct each other to a substantial degree. By analogy, the role of a defence lawyer in court is not to take an impartial or objective stance towards the defendant, but to make as strong an argument as possible for their innocence. It is not a problem that this induces biased thinking, because they cancel out with the opposite biases of the prosecuting attorney, and engaging the two in an iterative process of conversation and criticism is seen as a reliable (even if not perfect) path towards an accurate verdict.

We think that a practical implementation of the dialectical method could look as follows:

\begin{itemize}
    \item \textbf{Step 0 (pre-stage).} An organisation who sets up a team to work on a safety case ensures that team members are drawn from sufficiently diverse professional backgrounds and bring a broad range of expertise and perspectives. For our example, such a team could be expected to include at least cybersecurity experts, safety engineers and experts in frontier AI. 
    \item \textbf{Step 1 (team-internal).} Team members hold regular internal red-teaming sessions in which they identify potential weaknesses of their argument and register them as defeaters. Useful aids include automatic defeater searches with LLMs \cite{gohar2024codefeater}, pre-mortems (systematically considering the question “If our safety case failed, what would be the most likely reason why?”) and defeater checklists (see Section 5.1). 
    \item \textbf{Step 2 (organisation-internal).} The organisation appoints one or more internal challengers: subject-matter experts inside the organisation who are not directly involved in building the safety case, and who critically search the safety case for further defeaters, beyond those identified (and potentially addressed) before.
    \item \textbf{Step 3 (external).} The same as the previous stage, but with reviews conducted by third parties, such as evals organisations, government agencies or independent researchers (e.g. in the context of bug-bounty programs). 
\end{itemize}

After each of Steps 1-3, the safety case is adjusted in order to address the defeaters identified at that step, progressively strengthening the overall argument. Defeaters that cannot be readily addressed in this way will be registered as residual doubts to be quantified in the third and final stage of safety case development under Assurance 2.0 (see \textbf{Section \ref{sec6}}).

Practitioner reports suggest that the dialectical method, especially external reviews, can be helpful in identifying and addressing defeaters. For example, an external review of a Missile Defense system uncovered that it was vulnerable to a risk of inadvertent launch \cite{leveson2011safety}. According to someone involved in red-teaming the safety case, this discovery was due, among other things, to the different mindset with which evaluators approached the system, specifically the absence of a strong bias towards confirming the system’s safety. Moreover, in a recent expert survey on effective risk mitigation for general-purpose AI, some experts noted that safety cases require thorough external evaluations, as they might otherwise turn into a mere “box-checking exercise” \cite{uuk2024effective}.

By using the dialectical method, practitioners can identify defeaters and thereby work towards an ideal implicit in eliminative argumentation: the idea of strengthening an argument by considering relevant doubts and addressing them \cite{goodenough2015eliminative}. One question that we remain agnostic about in this context is whether eliminating a defeater should merely restore the initial confidence in the safety case (i.e. that existed before identifying the defeater), or whether successfully addressing a source of doubt should actually increase overall confidence. The question of how defeater resolution should impact overall confidence is an open question for further research.

\subsection{Defeater checklist: cyber misuse inability arguments for frontier AI}

In the table below a checklist is provided, which could be used by AI developers when creating safety cases for frontier AI. It provides a listing of different categories of defeaters and provides some examples, in the context of cyber misuse, for how such defeaters might arise, be mitigated or avoided.

\begingroup
\renewcommand{\arraystretch}{1.65}  
\begin{longtable}{|C{5.2cm}|C{7cm}|C{0.5cm}|}
\caption{Defeater checklist--example case of a frontier AI inability argument (cyber misuse harm)}\label{tab:defeaters_extended}\\
\hline
\rowcolor{gray!30} 
\textbf{Defeater} & \textbf{Example} & \textbf{\faCheck} \\
\hline
\endfirsthead

\hline
\rowcolor{gray!30} 
\textbf{Defeater} & \textbf{Example} & \textbf{\faCheck} \\
\hline
\endhead

\multicolumn{3}{|c|}{\cellcolor{orange!30}\textbf{Fallacious Reasoning}} \\
\hline

\textbf{Invalidity:} Have you ensured that all argument steps follow basic and intuitive rules of deductive inference? & \textit{You have added several ‘completeness claims’ to make explicit where an argument step depends on the assumption that all relevant cases have been identified.} &  \\
\hline

\textbf{Circularity:} Have you ensured that no node in the safety case explicitly or implicitly relies on the conclusion that it is trying to support? & \textit{You notice that an evaluation cited to support an inability claim was itself constructed under the assumption that AI systems lack the ability in question.} &  \\
\hline

\textbf{Equivocation:} Have you clearly defined key terms and ensured that they are used consistently throughout the safety case? & \textit{You notice that “uplift” has a more stringent meaning in claims higher up in the argument tree than in supporting claims further down.} &  \\
\hline

\textbf{Planning Fallacy:} Have you allocated sufficient time and resources, based on a realistic reference class, to fill any remaining gaps in the safety case? & \textit{After consultation with safety engineers in a different company, you double your estimate of how long it will take to validate a scoring rule.} &  \\
\hline

\multicolumn{3}{|c|}{\cellcolor{purple!30}\textbf{Future-related Uncertainty}} \\
\hline

\textbf{Unknown Capabilities of Threat Actors:} Have you accounted for evolving threat actor capabilities, particularly in adversarial or geopolitical contexts? & \textit{You consider the risk of a sophisticated actor employing undisclosed AI tools, such as advanced cyberwarfare mechanisms, to bypass current safety assumptions.} &  \\
\hline

\textbf{Unpredictable End-use Applications:} Have you assessed potential misuse scenarios beyond intended purposes, especially for general-purpose AI? & \textit{You include risks of AI-generated disinformation or adaptive misuse of open-weight models.} &  \\
\hline

\textbf{Unknown (ab)user Profiles:} Have you considered the diversity of potential malicious actors, including insiders, organized groups, or low-resource users? & \textit{You model scenarios involving insider threats, organized cybercriminals, and low-skill but persistent attackers.} &  \\
\hline

\textbf{Known Unknowns:} Have you accounted for interactions between AI systems and other technologies, including complex, cascading effects? & \textit{You assess dependencies on critical infrastructure and model cascading failures triggered by subtle AI malfunctions.} &  \\
\hline

\textbf{Black Swan Events:} Have you mitigated potential Black Swan Events that could disrupt system safety? & \textit{You perform extensive adversarial stress tests to uncover failure points under low-probability but high-impact conditions.} &  \\
\hline

\textbf{Unknown Unknowns:} Have you acknowledged that some risks remain inherently unpredictable? & \textit{You highlight the limitations of current evidence and formalize an adaptive monitoring plan for unforeseen risks.} &  \\
\hline

\multicolumn{3}{|c|}{\cellcolor{blue!20}\textbf{Completeness Uncertainty}} \\
\hline

\textbf{Interdependencies with Other Systems:} Have you assessed risks arising from interactions between the AI system and its surrounding infrastructure? & \textit{You consider cascading risks from vulnerabilities in external cybersecurity components affecting your system’s safety.} &  \\
\hline

\textbf{Unforeseen Ripple Effects:} Have you tested for unintended consequences stemming from system optimizations or emergent behaviors? & \textit{You conduct scenario analyses to uncover risks where a local optimization destabilizes a broader system.} &  \\
\hline

\textbf{Safety Case Interaction:} Have you ensured assumptions in individual safety cases are compatible with each other? & \textit{You cross-check assumptions between AI-specific and infrastructure-related safety cases to identify overlooked incompatibilities.} &  \\
\hline

\multicolumn{3}{|c|}{\cellcolor{green!30}\textbf{Cognitive Biases}} \\
\hline

\textbf{Confirmation:} Have you sought out relevant evidence against the safety claim? Have you sought alternative interpretations of evidence seemingly in support of the safety claim? & \textit{You notice that the proxy task you initially picked to evaluate a system’s capabilities is particularly difficult, and replace it by an easier one.} &  \\
\hline

\textbf{Overconfidence:} Have you critically assessed the reliability and completeness of your evaluation methods, e.g., by comparing them to independent baselines? & \textit{You notice that comparable evaluation methods have been shown to underelicit a capability and therefore re-assess your confidence in your own method.} &  \\
\hline

\textbf{Anchoring:} Have you reviewed initial assessments and data to ensure they do not disproportionately influence subsequent evaluations and decisions? & \textit{You discount an evaluation method for being insufficiently reliable in a high-stakes context, even if the method is no less reliable than popular alternatives.} &  \\
\hline

\textbf{Availability Heuristic:} Have you actively considered risk sources, risk models, evaluation methods, failure modes, etc., beyond those that immediately come to mind? & \textit{You include at least one threat vector in the safety case that nobody is talking about in the literature.} &  \\
\hline

\textbf{Authority:} Have you included judgments from lesser-known experts or organizations in your evidence? & \textit{You decrease your confidence in a fine-tuning technique to elicit hidden model capabilities, in response to a barely cited, but well-argued research paper.} &  \\
\hline

\textbf{Groupthink:} Has a team-internal ‘challenger’ been constantly screening your arguments for weaknesses? & \textit{After a challenger points out how limited the available evidence is on novel cyberattacks, you substantially decrease your confidence in the corresponding part of the tree.} &  \\
\hline

\end{longtable}
\endgroup

\subsection{How investigation of defeaters affects the assurance argument}

\begin{figure}[htbp]
  \centering
  \includegraphics[width=0.8\textwidth]{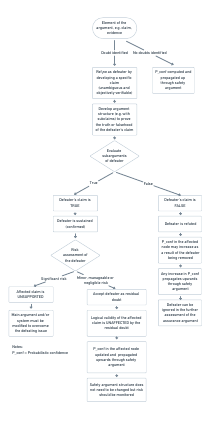}
  \caption{Flow chart showing the steps related to the handling of exploratory defeaters.}
  \label{fig:flowchart}
\end{figure}

The flow-chart in Figure \ref{fig:flowchart} summarises the process by which investigation of defeaters affects the assurance argument.

Figure \ref{fig:flowchart} is broadly self-explanatory, but it’s worth highlighting the following points (more details are available in \cite{bloomfield2022assessing}):

\begin{itemize}
    \item For defeaters classed as exploratory that are sustained, the logical validity of the affected node will be classed as \textit{UNSUPPORTED}. For those classed as exact, the logical validity of the affected node will become \textit{FALSE}.
    \item Through propagation, when a claim is \textit{UNSUPPORTED} all nodes further up the argument structure all the way to (and including) the top level claim will be \textit{UNSUPPORTED}. 
    \item If investigation of a defeater does not allow it to be refuted or sustained, then the argument node to which it points will remain \textit{UNSUPPORTED}.
\end{itemize}

\section{Confidence in the handling of residual risks}\label{sec6}

Bloomfield and Rushby identified three core categories of residual doubt in assurance cases, which we reproduce and summarise here for convenience:
\begin{enumerate}
    \item \textbf{Deductiveness Doubt:} Concerns about the logical coherence of reasoning within the assurance case.
    \item \textbf{Evidential Doubt:} Doubts about the quality, sufficiency, and reliability of evidence used to support claims. 
    \begin{itemize}
        \item Since great attention will have already been paid to confidence in evidence through the process for ‘weighing evidence’, there is a more narrow definition of what forms an evidential doubt might include. Bloomfield and Rushby give the example of minor systematic concerns which might apply to multiple items of evidence, and for which therefore the risk associated with the cumulative effect should be managed. 
    \end{itemize}
    \item \textbf{Interior Doubt:} Doubts about reasoning steps that are deductive and logically valid, but which are not sound given the context of use.
\end{enumerate}

We conjecture, though our analysis is preliminary, that the above set may not fully encompass the range of uncertainties that arise in the context of advanced AI safety cases, particularly when considering external factors and future developments. To address this gap, it may prove useful to introduce a new category:

\begin{enumerate}
    \setcounter{enumi}{3} 
    \item \textbf{Contextual Doubt:} Doubts arising from external factors and environmental uncertainties that are beyond the system's/developer’s control but significantly impact its safety.
\end{enumerate}

This latter type of doubt can be thought of as related to the concept of Operational Design Domain (ODD) - the specific environment and conditions where a system is designed to operate safely, and under which the safety case argument is conditioned upon. While Assurance 2.0’s original categories focus on the assurance case itself, Contextual Doubt addresses external factors that define the boundaries of reliable operation, or assumptions that are made about the operational environment when producing the safety case. 

Note that one way of managing doubts about the post-deployment environment is through in-operation monitoring of the safety case \cite{carlan2024dynamic}. 

\textbf{\underline{Application of Assurance 2.0 method to the cyber mis-use safety case:}}

The first step in addressing defeaters is to try and eliminate or mitigate them. The cyber mis-use safety case template \cite{goemans2024safety} identifies 8 defeaters and indicates that there is potential for those defeaters to be mitigated. For many of the defeaters, the authors provide hints as to what those mitigations might entail. If it were indeed found to be the case that the defeaters could be completely mitigated or eliminated, then there would be no residual doubts remaining in the safety case. However, since in our work, we wish to explore the problem and process of confidence assessment, we take one of the defeaters, $D1$ in Figure \ref{fig:fig3}, which we already discussed in Section 3.2, and consider what would be entailed in the handling of a corresponding residual doubt. Recall that, for the sake of argument we made an assumption that the residual doubt was with the timeliness of the incident monitoring, as opposed to with the ability of the incident monitoring to detect the novel cyber-attack (if it were to be given enough time). 

According to \cite{bloomfield2022assessing} the risk associated with residual doubts needs to be explicitly assessed and accepted. Risk assessment includes consideration of both the severity of harm and its likelihood. Bloomfield and Rushby indicate that risk assessment may be possible for some doubts but not all. However, it appears that Assurance 2.0 does not provide a method by which this risk assessment should happen. Nevertheless, if we map the approach to $D1$, then we can see that \textit{C2.2.1.1} acts in conjunction with \textit{C2.2.1.2} and \textit{C2.2.1.3} in support of claim \textit{C2.2.1} “Maximum time to detect AI enabled novel attack type and take frontier AI offline is 7 days”, and that failure to support \textit{C2.2.1} will in turn ultimately undermine the top-level claim. The requirement is then to determine the risk associated with a novel cyber attack causing harm, due to an inability to detect the novel cyber-attack within 1 day. In assessing risk, experts would need to consider both the level of potential harm caused by the novel cyber-attack and the likelihood of that attack occurring without it being detected in one day. In discussing these novel cyber-attacks Goemans et al state that since novel forms of AI enabled attacks are by definition not well understood, then it’s not possible to evaluate the system’s ability to conduct such hypothetical attacks, and it is for this reason that they proposed the solution of monitoring novel attacks post-deployment. But there is a problem here for us. We have assumed that there is a residual doubt against the incident monitoring solution, which according to Assurance 2.0 requires a risk assessment to be conducted, whilst at the same time, the very novelty of the attack means that it may not be possible to accurately foresee the specific hazards and hence quantify the specific risks. This conundrum may well be an issue that applies more generally in frontier AI, with relevance across different types of harm and different types of argument. Going forward, we can expect to be confronted with novel and unexpected behaviours and capabilities associated with frontier AI. An inability to be confident a priori that all novel hazards and harms have been mitigated, may be addressed by a ‘solution’ which involves the use of post-deployment monitoring, but that post-deployment monitoring can itself often be expected to have residual doubts associated with it, since it may not be possible to be confident that one can successfully monitor for things that are novel and unknown. The residual doubt in turn requires a risk assessment, which is difficult to perform convincingly because one is dealing with hazards and harms that are a priori unknown. One potential direction for handling this challenge of managing risk with a priori unknown situations is to place more emphasis on resiliency, with an argument being developed that such resiliency will be sufficient.

\section{Probabilistic valuation of confidence}\label{sec7}

As discussed in \textbf{Section \ref{sec2}}, a quantified and graduated level of confidence in a top-level claim can be useful for a number of purposes, including determining whether additional safety assurance work is necessary.

Assurance 2.0 supports graduated confidence analysis through a process called ‘probabilistic valuation’. The first component of probabilistic valuation involves determining the confidence in lower-level, evidence-supported leaf claims. These confidences are then recursively propagated up the argument structure of the safety case to give a confidence for the top-level claim. 

In subsection \ref{sec7_1}, we take Assurance 2.0’s process for determining the overall probabilistic valuation of belief in the top-level claim and apply it to the cyber misuse safety case. In subsections \ref{sec7_2} and \ref{sec7_3}, we outline some of the benefits and limitations of this process, respectively.

\subsection{Propagating confidence}\label{sec7_1}

To begin illustrating the propagation process, consider the expanded fragment of the cyber inability argument in Figure \ref{fig:fig3}, which we will refer to as the ‘\textit{C2.2.1 fragment}’.

We begin probabilistic valuation from the leaf claims because their associated confidence levels depend only on empirical evidence, not upon other claims. Accordingly, the leaf claims sit directly above evidence incorporation blocks in a safety case. It follows that \textit{C2.2.1.1}, \textit{C2.2.1.2}, and \textit{C2.2.1.3} are the leaf claims in the \textit{C2.2.1 fragment}. For each such claim \textit{C} and its supporting evidence \textit{E}, we assign a numerical posterior probability $P(C|E)$. By way of example, in Section 4, we have suggested a promising method based on the Delphi approach to obtain such posterior probabilities.

For the sake of illustration, henceforth we shall assume the following probabilities: 
$P(C2.2.1.1|E2.2.1.1) = 0.6$, $P(C2.2.1.2|E2.2.1.2) = 0.9$, and $P(C2.2.1.3|E2.2.1.3) = 0.8$. In Figure \ref{fig:fig3}, these posterior probabilities are listed on the lines between the Evidence Incorporation blocks and the evidence nodes below them.

In the next stage of the probabilistic valuation process, confidence values of lower-level claims are propagated upwards through the argument structure using a propagation method to calculate confidences of higher-level claims. We start by calculating confidence in leaf claims and then recursively calculate the confidence in each higher-level claim up to the top-level claim.

When calculating the confidence of a leaf claim, we consider the evidence and side-claim that support it. For example, in the \textit{C2.2.1 fragment}, the confidence in \textit{C2.2.1.1} depends on \textit{E2.2.1.1} and \textit{W2.2.1.1}. Continuing up the argument structure, the confidence in each higher-level claim depends on the sub-claims and side-claims one step below in the safety case. In the fragment, \textit{C2.2.1} is directly supported by \textit{W2.2.1}, \textit{C2.2.1.1}, \textit{C2.2.1.2}, and \textit{C2.2.1.3}. As we have already verified soundness before propagation, we make a stronger statement: \textit{W2.2.1}, \textit{C2.2.1.1}, \textit{C2.2.1.2}, and \textit{C2.2.1.3} in conjunction entail \textit{C2.2.1} \cite{bloomfield2022assessing}. Therefore,
\begin{equation}
    P(C2.2.1) = P(W2.2.1 \wedge \textit{C2.2.1.1} \wedge C2.2.1.2 \wedge C2.2.1.3)
\end{equation}

To generalise this, for claim $C$ whose decomposition into sub-claims $S_1, ..., S_n$ is supported by side-claim $W$, we would have

\begin{equation}
    P(C) = P(W \wedge S_1 \wedge \dots \wedge S_n)
\end{equation}
	
Assurance 2.0 proposes two propagation methods in particular that we can derive from this equation: the sum of doubts method and the product method \cite{bloomfield2022assessing}.

If we assume that the side-claim and all the subclaims are mutually independent, we can simply multiply them:

\begin{equation}
	P(C) = P(W) \times P(S_1) \times \dots \times P(S_n)
\end{equation}

This arithmetic product is the equation used in the product method of propagation.

To derive the expression for the sum of doubts method, we consider first the negation of $C$,

\begin{equation}
    P(\neg C) = P(\neg (W \land S_1 \land \dots \land S_n)) \\
= P(\neg W \lor \neg S_1 \lor \dots \lor \neg S_n)
\end{equation}

If all the negations on the right are mutually independent, the expression is equivalent to their sum. Otherwise, it will be less than the sum to the extent of their dependence. So,

\begin{equation}
    \begin{aligned} P(\neg C) & \leq P(\neg W)+P\left(\neg S_1\right)+\ldots + P\left(\neg S_n\right) \\ P(C)=1-P(\neg C) & \geq 1-\left(P(\neg W)+P\left(\neg S_1\right)+\ldots + P\left(\neg S_n\right)\right) \\ & =1-\left([1-P(W)]+\left[1-P\left(S_1\right)\right]+\ldots+\left[1-P\left(S_n\right)\right]\right) \\ & =1-\left(n+1-P(W)-P\left(S_1\right)-\ldots-P\left(S_n\right)\right) \\ & =1-n-1+P(W)+P\left(S_1\right)+\ldots+P\left(S_n\right) \\ & =P(W)+P\left(S_1\right)+\ldots+P\left(S_n\right)-n\end{aligned}
\end{equation}

This sum of doubts method uses the expression on the right as a lower bound on the confidence of \textit{C}. In essence, the sum of doubts method considers the probabilistic doubt of a claim (the probability of the negation of a claim, or 1 minus the probability of the claim) to be less than or equal to the sum of doubts in its sub-claims and side-claim \cite{bloomfield2022assessing}. As it provides a lower bound, it is highly conservative.

As we have noted, there are several assumptions underlying these methods. Both methods assume the safety case is sound and that each claim is entailed by the conjunction of its direct sub-claims and side-claim. Therefore, the fulfillment of all these claims is assumed to be sufficient and necessary. The product method and the lower-bound expression of the sum of doubts method assume mutual independence between claims. These assumptions simplify analysis, but they may not always hold. We examine these assumptions in subsections \ref{sec7_2} and \ref{sec7_3}.

The expressions above for these two propagation methods can be extended to the other types of CAE blocks by replacing the sub-claim probabilities with a posterior probability given evidence or a probability of a substituted sub-claim. We make these expressions explicit for each of the different blocks in the safety case in Table \ref{tab:sum_of_doubts} below:

\begingroup
\renewcommand{\arraystretch}{1.75}  
\begin{table}[ht]
\centering
\resizebox{\textwidth}{!}{
\begin{tabular}{|c|c|c|}
\hline
\textbf{Block} & \textbf{Sum of Doubts} & \textbf{Product} \\
\hline
Evidence Incorporation & 
$P^{D}_{\text{conf}}(C) \geq P(C \mid E) + P_{\text{conf}}(W) - 1$ & 
$P^{P}_{\text{conf}}(C) = P(C \mid E) \times P_{\text{conf}}(W)$ \\
\hline
Substitution & 
$P^{D}_{\text{conf}}(C) \geq P_{\text{conf}}(W) + P_{\text{conf}}(S) - 1$ & 
$P^{P}_{\text{conf}}(C) = P_{\text{conf}}(W) \times P_{\text{conf}}(S)$ \\
\hline
Decomposition & 
$P^{D}_{\text{conf}}(C) \geq P^{D}_{\text{conf}}(W) + \sum_{i=1}^{n} P^{D}_{\text{conf}}(S_i) - n$ & 
$P^{P}_{\text{conf}}(C) = P^{P}_{\text{conf}}(W) \times \prod_{i=1}^{n} P^{P}_{\text{conf}}(S_i)$ \\
\hline
\end{tabular}}
\vspace{0.3cm}
\caption{$P^{D}_{c}(C)$ and $P^{P}_{c}(C)$ represent the probabilistic confidence of claim \textit{C} calculated with the sum of doubts and product methods, respectively. All arguments on the right sides of the equations and inequalities refer to sub-claims \textit{S} and side-claims \textit{W} that immediately support \textit{C}. The expressions are taken from Assurance 2.0 \cite{bloomfield2022assessing}}
\label{tab:sum_of_doubts}
\end{table}
\endgroup

For both these methods, we must assign probabilistic confidences to each side-claim based on its strength. Additionally, although not explicitly appearing in the equations above, Assurance 2.0 allows us to manually adjust the propagation with a factor \textit{f} or to manually override individual calculations for specific circumstances should it be deemed appropriate, e.g. to account for significant residual doubts. 

Executing this process by assigning probabilities and propagating through the entire argument will give an overall probabilistic confidence for the top-level claim. To illustrate these propagation methods in practice, we will now apply each method without manual adjustments to calculate the confidence of \textit{C2.2.1} in the fragment using the probabilities given in Figure \ref{fig:fig3}. In Figure \ref{fig:fig7} below, we show probabilistic confidence calculations propagating up with the sum of doubts method:

\begin{figure}[htbp]
  \centering
  \includegraphics[width=1\textwidth]{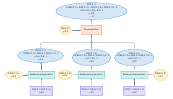}
  \caption{Probabilistic confidence propagated up the C2.2.1 fragment of Figure \ref{fig:fig3} using the sum of doubts method.}
  \label{fig:fig7}
\end{figure}

Note that the result of the sum of doubts method in this instance produces a negative result. This will happen whenever the total sum of doubts across the safety case is greater than $1$. As sum of doubts provides a lower bound and negative probabilities are not meaningful, we simply round such results up to $0$. Therefore, the sum of doubts method suggests a confidence of $0$ for \textit{C2.2.1}, given the example probabilities that we used.

\begin{figure}[htbp]
  \centering
  \includegraphics[width=1\textwidth]{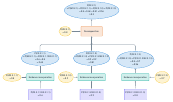}
  \caption{Probabilistic confidence propagated up the C2.2.1 fragment of Figure \ref{fig:fig3} using the product method}
  \label{fig:fig8}
\end{figure}

In Figure \ref{fig:fig8}, we show the propagation calculations using the product method instead.

The product method produces a more optimistic confidence of $0.2$ for \textit{C2.2.1}.

We see that the sum of doubts method produces a confidence of zero, whilst the product method  produces a small positive overall confidence. If a decision-maker were depending on a good probabilistic valuation to make a deployment decision they would clearly want a significantly higher level of confidence. To achieve greater confidence in the top-level claim, the confidence in leaf claims and side claims must also be considerably greater.

For example, consider the required confidence for side claims and posterior probabilities given evidence (the probabilities that are assigned or measured rather than calculated through propagation) should we designate a threshold of $95\%$ overall confidence as sufficient for the \textit{C2.2.1} fragment. To achieve this overall confidence, individual assigned probabilities cannot go below $0.95$ because this probability will only decrease when summed with doubts of other claims or multiplied with other probabilities less than 1. However, if any one assigned probability is $0.95$, all the other assigned probabilities must then be 1 to keep the overall confidence at $95\%$, which, practically speaking, is impossible for our particular safety case fragment. Instead, to minimise the greatest assigned probability, we shall consider the required level of confidence when all assigned probabilities are equal.

If we assign each probability a value of \( p \), for the sum of doubts:
\begin{align*}
P(C2.2.1) &= P(W2.2.1) + P(C2.2.1.1 \mid E2.2.1.1) + P(W2.2.1.1) \\
&\quad + P(C2.2.1.2 \mid E2.2.1.2) + P(W2.2.1.2) \\
&\quad + P(C2.2.1.3 \mid E2.2.1.3) + P(W2.2.1.3) - 6
\end{align*}

\[
0.95 = 7p - 6
\]
\[
0.99286 = p
\]

Meanwhile, for the product method:
\begin{align*}
P(C2.2.1) &= P(W2.2.1) \times \big[P(C2.2.1.1 \mid E2.2.1.1) \times P(W2.2.1.1)\big] \\
&\quad \times \big[P(C2.2.1.2 \mid E2.2.1.2) \times P(W2.2.1.2)\big] \\
&\quad \times \big[P(C2.2.1.3 \mid E2.2.1.3) \times P(W2.2.1.3)\big]
\end{align*}

\[
0.95 = p^7
\]
\[
0.99270 = p
\]

For both methods, the confidence for assigned probabilities must be around $99.3\%$ to achieve the desired threshold. Given the many complex sociotechnical factors involved in cyber inability and the gaps in our understanding of frontier AI systems, this very high level of confidence will be extremely difficult to achieve for any of the claims in the fragment. Unfortunately, the confidence levels required in practice would have to be even greater to achieve a $95\%$ confidence in a top-level claim, since the fragment we have considered is only part of a much larger safety case. In addition to which, even a confidence of $95\%$ in the top-level claim may easily be insufficient for high risk safety cases.

Additionally, note that the discrepancy between the two methods is much smaller in this analysis than when we were propagating up with lower confidences. As such, the results of the two methods tend to be much closer for higher-confidence arguments \cite{bloomfield2022assessing}.

The exercises above highlight how quickly confidence falls and how difficult it is to obtain high overall confidences for a safety case when propagating with these methods. Bloomfield and Rushby make the same observations and conclude that, in practice, it may often not be viable to achieve reliable absolute probabilistic valuations. 

We have already identified that the sum of doubts method produces conservative lower bounds results, but we see that the product method can also produce strikingly low confidences. We consider why this might be in subsection \ref{sec7_3}.

\subsection{Benefits of probabilistic assessment with Assurance 2.0}\label{sec7_2}

A key strength of Assurance 2.0 is that the process of assessing soundness is explicitly separate from and a prerequisite to probabilistic valuation \cite{bloomfield2022assessing}. As a result, the probabilistic valuation benefits from a solid foundational structure of well-defined building blocks that has withstood deliberate scrutiny of its evidence and reasoning. Without a separate analysis of soundness, other probabilistic approaches assume soundness or attempt to verify it when evaluating confidence \cite{bloomfield2022assessing}. The results of such approaches are questionable because probabilistic methods themselves are not well-suited to validating the logical foundations of the safety case \cite{bloomfield2022assessing}.

In addition, both the sum of doubts and product methods are much simpler than other common propagation methods like BBNs and DST. They require less information to implement and less computation to execute. Involving only basic arithmetic, both methods are straightforward enough for non-technical decision-makers to grasp and even verify manually. Through greater transparency, understanding, and trust, these approaches can more effectively communicate risk to a broader audience. Therefore, they may be the better choice for simple analyses of low-stakes scenarios with limited info, where complex approaches may overcomplicate analysis while ultimately providing limited extra value.

Designated software further simplifies the implementation of Assurance 2.0 probabilistic valuation. Clarissa, a software tool specifically designed to support the Assurance 2.0 methodologies, provides a structured environment for building and managing assurance cases, defining claims, linking claims through arguments, and attaching supporting evidence \cite{varadarajan2023clarissa}. This tool automatically enforces the core principles of Assurance 2.0, helping to ensure the validity and soundness of the logical arguments presented. Clarissa is equipped with a specialized plugin designed to calculate probabilistic confidence using the sum of doubts approach. Clarissa also facilitates manual adjustments to calculated probabilistic confidence values and automatic colour-coding of nodes based on their confidence. The latter feature offers a visual representation that allows assessors to quickly gain an understanding of confidence and uncertainty across the safety case.

Overall, Assurance 2.0 probabilistic valuation is a reasonable choice for confidence assessment when we wish to build on the foundation of soundness that Assurance 2.0 provides and only require a rough probabilistic analysis of confidence across the safety case. As the sum of doubts method is relatively pessimistic, it may be a good fit when looking for a conservative estimate of confidence for small, high-confidence safety arguments.

\subsection{Limitations of probabilistic assessment with Assurance 2.0}\label{sec7_3}

On the other hand, there are several limitations to Assurance 2.0’s probabilistic valuation. One key limitation is that the calculated probabilistic confidence for the top claim may not be a reliable indicator of overall system safety, often producing very low confidences \cite{bloomfield2022assessing}. As confidence values are propagated upwards through the argument structure, even minor doubts at lower levels can accumulate and significantly reduce the final confidence in the top claim. This is especially true for the sum of doubts method, where probabilistic doubt in any minor claim of the safety case linearly affects the confidence of the top-level claim. Without manual adjustments, confidence can quickly drop to 0 for bigger or lower-confidence arguments, precluding deeper analysis.

The assumptions that underlie Assurance 2.0’s probabilistic valuation simplify analysis, but they rule out more accurate, complex modelling of the relationships between claims in the cyber inability safety case. By assuming that sub-claims are mutually independent, that they aggregate to the higher-level claim through conjunction, and that they are equally important, we exclude the consideration of the dependencies between sub-claims, the possibility of disjunction, and the varying influence of sub-claims on the claim above. In the \textit{C2.2.1} fragment, consider \textit{C2.2.1.1} (“\textit{Incident monitoring detects all novel cyberattacks within 1 day}”), \textit{C2.2.1.2} (“\textit{Revision protocol ensures safety case is updated within 5 days of novel cyber attack being detected}”) and \textit{C2.2.1.3} (“\textit{The AI system will be taken offline within 1 day of top-level safety case claim becoming false}”). Although they are distinct claims about separate parts of the system, they are not fully independent. The claims might be correlated based on common influencing factors. For example, the same budget, team of people, and research insights may be shared across the functions of incident monitoring, safety case revision, and taking the AI system offline. Hence, if the safety case is updated more quickly, then additional time may be available to implement the emergency protocol and take the AI system offline before harm is done. Additionally, the decision of whether to even take the AI system offline would be dependent on the information available from incident monitoring. Furthermore, we might consider \textit{C2.2.1.1} more important to the higher-level claim \textit{C2.2.1} of “\textit{maximum time to detect AI enabled novel attack type and take the frontier AI offline is 7 days}” because, as a safety measure earlier in the process, it could allow planning of alternative courses of action. However, the propagation methods will not account for these dependencies and differentials in importance.

The assumptions in the propagation methods may also contribute to the low confidences produced by probabilistic valuation. As claims are assumed to be independent and necessary in conjunction, adding a new claim with any uncertainty to the safety case will always decrease the overall confidence. The new confidence of the top-level claim will be lesser by the degree of doubt in the new claim if using sum of doubts or will be multiplied by the new claim’s confidence if using the product method. Consequently, safety cases with more claims that may simply be more thorough will tend to have lower confidence levels. To deal with this issue, Bloomfield and Rushby suggest that some manual increase in confidence values, e.g. through use of the \textit{f} factor, may be used to account for the increased confidence that arises from taking a larger number of smaller, better-justified steps in the argument that may exist in a more rigorous case. 

Confidence in a case can be improved where an additional sub-claim can be found such that the two sub-claims jointly propagate a greater confidence than they would individually. This specific relationship might resemble that which we discussed in weighing evidence in Section 3.2. This could be achieved using a “confidence-building block” wherein an argument would be made that confidence from the 2 sources of evidence would lead to a relative increase in confidence in the parent claim on the basis that the sub-claims rely on sufficiently different underlying mechanisms. 

Although the probabilistic valuation process does allow some flexibility through adjustments, these adjustments are ad-hoc and subjective. As mentioned earlier, Assurance 2.0 allows adjustments to confidence values using factors to account for the strength of justifications and even permits manual overrides of calculated values for individual blocks \cite{bloomfield2022assessing}. Adjustments of this nature allow expert judgment and context-specific considerations into the assessment, but they also introduce subjectivity that complicates reproducibility and comparisons between different assessments. To make these subjective decisions more structured rather than arbitrary, we suggest defining more explicit and transparent processes akin to what was achieved with the LLM-based Delphi method above, placing specific constraints upon adjustments, or at least supporting subjective decisions with justification.

When these propagation methods are inaccurate or their assumptions do not hold, we may prefer a different propagation method altogether. Although Assurance 2.0 proposes the sum of doubts and product propagation methods in particular, it accepts other propagation methods as long as they are used consistently \cite{bloomfield2022assessing}. BBNs and DST are more sophisticated propagation methods that allow us to overcome some of the limitations of the sum of doubts and product methods. BBNs allow conditional probabilities to give differential importance across sub-claims and can employ noisy operators, such as noisy-OR and noisy-AND, to model more complex relationships than simple conjunction. DST is even more flexible and allows us to set probability masses, i.e. degrees of belief, to sets of claims rather than just single ones, to perform disjunctive belief calculations, to separate uncertainty from disbelief, and to establish multi-node arguments for nodes that can be complementary or redundant. If we can make effective use of the greater flexibility of these methods, they may better assess confidence in our safety case. Nonetheless, we should still assess the soundness of the safety case separately before applying these probabilistic methods.

In summary, Bloomfield and Rushby \cite{bloomfield2022assessing} state that probabilistic valuation should primarily be used with less critical systems. They also state that absolute probabilistic valuations should be used with caution, both on account of the conservatism in Assurance 2.0’s recommended probabilistic propagation mechanisms and due to the subjectivity in assessing probabilistic confidence in evidence and assumptions.

\section{Resolution of defeaters}\label{sec8}

\subsection{The role of defeaters in Assurance 2.0 safety arguments}

As part of the dialectic method described in \textbf{Section \ref{sec5}} above, it behooves the safety case team and internal or external reviewers to identify all defeaters and for each, develop an argument structure that can be used to prove its truth or falsehood. The team then gathers evidence that can be incorporated into the argument structure to either:

\begin{enumerate}
    \item \textbf{Refute the defeater} in which case it need no longer be considered in the assessment of the assurance argument

or

\item \textbf{Sustain the defeater} in which case the main argument and or the AI system being assessed must be modified to overcome the defeating issue
\end{enumerate}

Until the defeater is either refuted or sustained, the claim it affects must be considered unsupported and following Assurance 2.0’s principles of propagation, every node above that all the way to the top-level safety claim will be unsupported (probabilistic confidence that was associated with these nodes will be meaningless, however, probabilistic confidence associated with other nodes in the argument may remain). This means that until ALL potential defeaters have been either dismissed or accepted as residual doubts, a safety case is not complete and the top level claim will not be supported. 

However the order in which defeaters are addressed can have a significant impact on both the time and amount of effort it takes to either increase confidence in the overall case, or to reach the conclusion that the overall argument is not supported and mitigation measures must be put in place such as modifications to the system or, if already deployed, taking it offline. In this latter case, by reducing the time to reach this decision, it may be possible to reduce the exposure time and therefore reduce the risk of harm. Whilst Assurance 2.0, when implemented rigorously would prevent a system from being deployed with unsupported defeaters that have not yet been investigated, the process will be more resilient if it can accommodate changes in context such as new threats or vulnerabilities being identified post-deployment and provide a route to identifying and investigating potential defeaters as rapidly as possible and addressing the most significant ones first. The rapid pace of change in frontier AI and the resource constraints and time pressures that safety teams must deal with, makes this prioritisation even more crucial, compared with slower moving and more established fields of safety engineering. 

This section explores the consequences of either sustaining or refuting a defeater then explores the different factors that would affect the optimal order in which to tackle defeaters in an assurance argument. We propose an approach to systematically prioritise efforts to resolve a number of defeaters in a safety case and then apply it to the safety case fragment of Figure \ref{fig:fig3} as an example.

\subsection{Factors affecting prioritisation of defeater assessment}

In Section 5.2 we summarized the process by which investigation of defeaters affects the assurance argument. In this section we discuss which factors are pertinent when prioritising the order with which defeaters should be assessed.

Resolution of defeaters is a process that will often need to be tackled sequentially. Each defeater has factors that contribute to the impact its resolution will have on the safety case argument. Some factors are determined by the defeater’s explicit definition whilst others are based on assessments of the probability that the defeater will be sustained or the expected amount of effort required for its resolution. There are also contextual factors such as the defeater’s location within the argument structure. This section outlines these factors, and the following section proposes a strategy for using these factors to prioritise the order in which to investigate defeaters.

This approach is based on an assumption of mutual independence between defeaters. Further discussion of the need for a method that handles defeaters with conditional probabilities follows in Section 8.6. 

\begingroup
\renewcommand{\arraystretch}{1.5}  
\begin{longtable}{|C{5cm}|C{7cm}|C{0.5cm}|}
\caption{Which defeater factors to consider when deciding on prioritisation}\label{tab:defeaters_factors}\\
\hline
\rowcolor{gray!30} 
\textbf{Factor} & \textbf{Prioritisation Considerations} \\
\hline
\endfirsthead

\hline
\rowcolor{gray!30} 
\textbf{Factor} & \textbf{Prioritisation Considerations}\\
\hline
\endhead

\multicolumn{2}{|c|}{\cellcolor{orange!30}\textbf{Defeater Characteristics}} \\
\hline

Prior estimate of the probability that the defeater would be sustained (confirmed) & Safety teams can estimate the probability that a defeater would be sustained as a value between 0-1. Investigate the defeaters that are most likely to be sustained before ones that are likely to be refuted or result in residual doubts so that expected time to uncover a valid defeater is as low as possible \\
\hline

Logical assessment of defeater & Only unsupported defeaters still need be assessed. Defeaters which have already been proven to be valid or false need not be investigated again.\\
\hline

How much effort would be required to either sustain or refute the defeater through its sub-argument structure (in relation to effort required to sustain or refute other defeaters in the argument)? & Tackle defeaters that would require lower effort to investigate first. Amount of effort may depend on availability of evidence, requirement for specific expertise, requirement for new tests or evaluations to be developed and conducted. By taking the 'low-hanging-fruit' approach, more defeaters can be assessed with a given amount of resource. We acknowledge that the 'indefeasibility' principle requires ALL defeaters to be resolved, however reducing the number as quickly as possible can increase the overall probabilistic confidence in the top level claim even if it remains logically unsupported, and may lead to identifying ‘show-stopper’ defeaters more quickly. \\
\hline

What kind of modifications would be required to the safety case argument or the system if the defeater is sustained? & Defeaters which, if sustained, would require more significant modifications to the assurance argument or to the system being assessed should be prioritised before those where modifications are localised, narrower in scope or would be easier to implement. This is because more significant changes may invalidate analysis conducted on the previously configured system / argument structure, rendering it invalid. By finding defeaters that require more significant modifications earlier, we reduce the amount and likelihood of rework/wasted effort.\\
\hline

\multicolumn{2}{|c|}{\cellcolor{orange!30}\textbf{Affected node characteristics}} \\
\hline

What type of node is affected by the defeater? & (Evidence / reasoning step / side-claim) Defeaters relating to reasoning steps may affect the logical validity of the safety case argument - these are therefore critical to resolve first.
Those relating to evidence may require empirical work (evaluation/testing) which depending on specifics may take more effort. The scope of side-claims may be broad, however defeaters affecting those relating to contextual assumptions may require major work to assess. \\
\hline

\multicolumn{2}{|c|}{\cellcolor{orange!30}\textbf{Argument structure}} \\
\hline

Impact on probabilistic confidence in top level claim if defeater is refuted & Calculate the confidence of the top level claim (using Assurance 2.0 propagation principles) before investigating the defeater. Then calculate the confidence assuming the defeater was refuted. Subtract the second confidence from the first to get a measure of the impact.
Investigate defeaters whose refutation would increase probabilistic confidence in the overall safety case by the greatest amount \\
\hline
Number of dependent claims affected (and how many of these dependent claims are affected by other defeaters) & In general the defeaters that are highest in the argument structure should be prioritised first. If they are sustained then the modification to the system or argument structure may render defeaters lower in the safety case structure irrelevant. Therefore had those defeaters that are lower in the argument been investigated first, the effort would have been wasted.\\
\hline

\multicolumn{2}{|c|}{\cellcolor{orange!30}\textbf{Contextual factors}} \\
\hline

Temporal availability of facilities (e.g. compute or testing infrastructure) or expertise & If facilities (compute or testing infrastructure) or experts are scarce and available only for a limited period, then investigating those defeaters that require them should be prioritised whenever the resources become available.\\
\hline
\end{longtable}
\endgroup

\subsection{Impact on probabilistic confidence of refuting defeaters}

If we assume the position that probabilistic confidence improves as defeaters are refuted, then  as described in Section 7, the updates to probabilistic confidence can be propagated upwards through an assurance argument either by ‘sum of doubts’ or ‘product’ methods. We will use the product method to illustrate how refuting defeaters can affect probabilistic confidence in parent claims.

Let us use the safety case fragment for \textit{C2.2.1} (Figure \ref{fig:fig3}). Below, we calculate the probabilistic confidence in the main claim $P_{conf}(C2.2.1)$ firstly without considering defeaters, using hypothetical values for \textit{P(C2.2.1.1 | E2.2.1.1)}, \textit{P(C2.2.1.2 | E2.2.1.2)} and \textit{P(C2.2.1.3 | E2.2.1.3)}. Our side-node claims (provenance of evidence) have also been ascribed hypothetical probabilistic confidences for the purposes of illustration.

We then estimate how much the probabilistic confidence in the claims directly affected by defeaters \textit{D1} and \textit{D2} would increase if these defeaters were refuted. We can then repeat the propagation calculations for each case to compare how much the top level claim is affected by the refutation of each defeater:

\begin{table}[ht]
\centering
\renewcommand{\arraystretch}{1.5}  
\resizebox{\textwidth}{!}{
\begin{tabular}{|l|c|c|c|}
\hline
\textbf{} & \textbf{Baseline - no defeaters} & \textbf{Defeater \textit{D1} refuted} & \textbf{Defeater \textit{D2} refuted} \\
\hline
\multicolumn{4}{|l|}{\textbf{Leaf node inputs}} \\
$P(C2.2.1.1 \mid E2.2.1.1)$ & 0.6 & 0.85$^*$ & 0.6 \\
$P(W2.2.1.1)$ & 0.8 & 0.8 & 0.8 \\
$P(C2.2.1.2 \mid E2.2.1.2)$ & 0.9 & 0.9 & 0.9 \\
$P(W2.2.1.2)$ & 0.9 & 0.9 & 0.9 \\
$P(C2.2.1.3 \mid E2.2.1.3)$ & 0.8 & 0.8 & 0.9$^*$ \\
$P(W2.2.1.3)$ & 0.7 & 0.7 & 0.7 \\
\hline
\multicolumn{4}{|l|}{\textbf{Calculated using 'Arithmetic Product' propagation method}} \\
$P_{\text{conf}}(C2.2.1.1)$ & 0.48 & 0.68 & 0.48 \\
$P_{\text{conf}}(C2.2.1.2)$ & 0.81 & 0.81 & 0.81 \\
$P_{\text{conf}}(C2.2.1.3)$ & 0.56 & 0.56 & 0.63 \\
$P(W2.2.1)$ & 0.8 & 0.8 & 0.8 \\
\textbf{$P_{\text{conf}}(C2.2.1)$} & \textbf{0.17} & \textbf{0.25} & \textbf{0.20} \\
\hline
\multicolumn{4}{|p{14cm}|}{\textbf{Comments:} Refuting \textit{D1} causes $P(C2.2.1.1 \mid E2.2.1.1)$ to increase. Through propagation, $P_{\text{conf}}(C2.2.1)$ increases by $0.08$ from $0.17$ to $0.25$. Refuting \textit{D2} causes $P(C2.2.1.3 \mid E2.2.1.3)$ to increase. Through propagation, $P_{\text{conf}}(C2.2.1)$ increases by $0.03$ from $0.17$ to $0.20$.} \\
\hline
\end{tabular}
}
\vspace{0.3cm}
\caption{Comparison of the impact of refuting \textit{D1} vs. \textit{D2} on the probabilistic confidence of the parent claim $C2.2.1$. Asterisks indicate hypothetical values for the updated probabilistic confidences assuming defeaters are refuted have been used for illustrative purposes.}
\label{tab:comparison_defeaters}
\end{table}

In this example, refuting \textit{D1} has a greater impact ($0.08$) on the confidence in \textit{C2.2.1} than refuting \textit{D2} (impact of $0.03$).

\subsection{Strategy for prioritisation of defeater resolution}

Following Assurance 2.0’s paradigm, that logical soundness is paramount, we propose the following strategy:

\begin{enumerate}
    \item \textbf{Step 1.} Look first at each defeater’s potential impact on logical soundness of the argument (e.g. if it challenges a reasoning step). All defeaters falling into this category should be tackled before those which do not.
    \item \textbf{Step 2.} Secondly, assess the type of modifications that would be required either to the system or to the assurance argument if the defeater were sustained. Defeaters where the modifications require restructuring the argument, or will be widest in scope of modifications required to the system should be tackled before those affecting a more localised part of the argument or with narrower scope.
    \item \textbf{Step 3.} Once all defeaters that affect logical validity have been addressed, a prioritisation score can be calculated for the remaining defeaters using a weighted scoring model approach based on these inputs:
    \begin{itemize}
        \item \textbf{Impact:} The impact on the probabilistic confidence of the top level claim of refuting defeater. (value in the range 0-1, calculated as described in Section 8.3)
        \item \textbf{Probability:} Prior estimate of the probability that the defeater would be sustained. (value in the range 0-1)
        \item \textbf{Effort:} How much effort is required to resolve the defeater. The safety team should make a subjective relative assessment. (value in the range 0-1)
    \end{itemize}

\end{enumerate}

The expected ‘benefit’ of tackling a defeater depends on the prior probability that it will be sustained and the impact that refuting it would have on the top level claim. The ‘cost’ of resolving it depends on the effort involved.

A number of factors will affect the relative significance of each of these factors and they will likely vary from one safety case and team to another. For example a team with significant resource constraints may be more sensitive to \textit{Effort} than a team with more abundant resource availability. We therefore propose a weight is assigned to each of the inputs. These weights would be set for a safety case ahead of the prioritisation activity, where the weights are $W_{Probability}$,$W_{Impact}$ and $W_{Effort}$:

\begin{equation}
    \text{Prioritisation score} =  \frac{(W_{Probability} \times \text{Probability}) + (W_{Impact} \times \text{Impact})}{(W_{Effort} \times \text{Effort})}
\end{equation}

For defeaters that do not impact the logical soundness of the argument and would not require it to be restructured if sustained, it would be preferable to investigate those with higher prioritisation scores before those with lower scores.

This approach is limited to handling defeaters that are mutually independent and would need to be extended to handle defeaters with conditional probabilities - see further discussion of this in section 8.6

\subsection{Example application to safety case fragment}

We take the safety case fragment of Figure \ref{fig:fig3}  as an example to illustrate how prioritisation can be assessed. 
Assuming the safety case team has to decide whether to prioritise the investigation of \textit{D1} or \textit{D2}, we end up with Table 6. 

\begin{table}[ht]
\centering
\renewcommand{\arraystretch}{1.5} 
\begin{tabular}{|p{4.5cm}|p{4cm}|p{4cm}|}
\hline
\textbf{Factor} & \textbf{\textit{D1}} & \textbf{\textit{D2}} \\
\hline
\textbf{} & Historical evidence of non-AI enabled novel cyber-attacks being launched and remaining undetected for many weeks & Evidence of shutdown protocol not being followed due to pressure on decision makers to keep the system online. \\
\hline
\textbf{Probability:} Prior estimate of the probability that the defeater would be sustained (hypothetical values for illustrative purposes) & 0.75 & 0.65 \\
\hline
\textbf{Effort:} How much effort would be required to either sustain or refute the defeater? (subjective assessment) & 0.6 & 0.8 \\
\hline
\textbf{What kind of modifications would be required to the safety case argument or the system if the defeater is sustained?} & The system could be modified so that its ability to cause novel cyberattacks was reduced, and then new evidence collected and evaluated. & The system’s context could be modified to introduce independent authority to oversee shutdown decision-making. \\
\hline
\textbf{What type of node is affected by the defeater?} & Evidence & Side claim \\
\hline
\textbf{Impact:} How much does the probabilistic confidence in the top level claim increase if the defeater is refuted? & 0.08 & 0.03 \\
\hline
\end{tabular}
\vspace{0.3cm}
\caption{Prioritisation of the impacts of resolving \textit{D1} vs. \textit{D2}}
\label{tab:defeater_comparison}
\end{table}

Before assessing probabilistic confidence, first we must assess whether either defeater challenges a reasoning step or may require wider scope changes to the system or restructuring of the argument (Steps 1 and 2 in Section 8.4). In this case, on the basis that a reasoning step is not affected and that neither of the required modifications would be significantly more challenging, we proceed to Step 3 (Note that a rigorous analysis of this safety case example fragment may find other modifications that would affect the structure of the argument and have implications for the prioritisation). 

Assuming weighting factors of 1, we calculate the priority scores using the above formula to be:

\begin{itemize}
    \item Prioritisation score for \textit{D1}:
    \begin{equation}
        \frac{(1 \times 0.75) + (1 \times 0.08)}{(1 \times 0.6)} = \textbf{1.38}
    \end{equation}
    \item Prioritisation score for \textit{D2}: 
    \begin{equation}
        \frac{(1 \times 0.65) + (1 \times 0.03)}{(1 \times 0.8)} = \textbf{0.85}
    \end{equation}
\end{itemize}

The defeater with the higher prioritisation should be addressed first. In this example, this is \textit{D1} with its prioritisation score of $1.38$ rather than \textit{D2} with its score of $0.85$. The weighting factors will have a significant effect on the outcomes - in this example, the absolute values of ‘impact’ are an order of magnitude lower than those for ‘probability’. The safety team should conduct a sensitivity analysis to inform the selection of weighting factor values.

\subsection{Discussion of limitations}

The limitations relating to subjectivity and the challenges with quantifying probabilities outlined in section \ref{sec7_3} also apply to the resolution of defeaters. 

We believe that as a result of the complexity of frontier AI systems and the emergent nature of their behaviour, defeaters may have inter-dependencies that also require a more sophisticated approach to avoid over-simplifying our assessment of their combinations.

Examples of the types of dependencies that may exist between defeaters:

\begin{itemize}
    \item \textbf{Probability:} \textit{D1} is more likely to be sustained if \textit{D2} is also sustained than if \textit{D2} is refuted. 
    \item \textbf{Effort:} Investigation of \textit{D1} involves running an evaluation that also provides evidence that would sustain or refute \textit{D2}.
    \item \textbf{Impact:} If \textit{D1} is sustained, the system or argument structure would need to be modified in a way that makes \textit{D2} no longer applicable.
    \item \textbf{Grouping:} \textit{D1}, \textit{D2} and \textit{D3} may form a composite group that should be assessed together rather than in sequence
\end{itemize}

We propose follow-up work, which could take the form of developing a safety case fragment involving inter-dependent defeaters and exploring and comparing different techniques to identify and prioritise defeaters. Acknowledging the findings of \cite{graydon2017investigation} which found failings in implementations of methods such as Dempster-Shafer Theory and BBNs in previous studies (discussed in Section 2.2), it will be necessary to ensure the methodology is applied meticulously before drawing conclusions. 

If a method is developed for combining probabilistic confidence in defeaters and claims with interdependencies, then the defeater prioritisation formula above will likely need to be updated to accommodate this. Where weighting factors are still involved (for the 3 variables used: impact, probability and effort), further research is required to develop a methodology for assigning the weights to each.

\section{Communicating confidence}\label{sec9}

There are at least two classes of target audience for the communication of confidence information. One class are the people with safety engineering responsibilities, employed either within the AI developer or within an external auditing function. The second class is an executive decision maker, either within an AI developer or employed by a regulator who needs to make a go / no-go decision on deployment of an AI model or allowance of its continued operation. In this section we focus our attention on how to communicate confidence to this second ‘executive’ class. 

Clear, concise and visual information is the most effective way to communicate confidence levels \cite{eppler2009systematic}. Visual information can help communicate safety case conclusions. Words and numbers can be used to clarify this visual information, including the level of uncertainty surrounding the confidence level and whether there is sufficient evidence \cite{trevena2013presenting}, \cite{hausmann2008sequential}. Decision-makers must also understand how outcomes may change over time, due to the shifting nature of AI development. Using cyber misuse as an example, decision-makers need to understand how the threat landscape of risks may shift, as the arms race of attack / defence intensifies.

For our paper, we assume a framework for rational decision-making that includes four stages \cite{spiegelhalter2017risk}: (1) structuring the list of actions and possible consequences available, (2) giving a value to those possible futures, (3) assigning a probability for each consequence given each action, and (4) establishing the rational decision which maximizes the expected benefit. Decision-makers should ideally be given information regarding stages 1-3 as part of the visual information. In the context of an executive making a decision on frontier AI given the potential of cyber misuse, the list of actions (1) might be (‘approve to deploy’, ‘do not approve to deploy’). Giving a value to each of those decisions (item 2) would necessitate a communication about both benefit and risk, where in this section we concern ourselves with just the communication of the risk element. (3) decision-makers require information to ascertain the probability of each consequence given each action, and this section addresses the need for qualitative (and if possible) quantitative means of reaching such probabilities for AI deployment. 

Be aware that decision-makers have their own biases, therefore it is important to consider the perceptions, emotions and feelings that visual information may invoke in the audience. 

\subsection{Tailoring communication for the target audience}

There is a need to build trust in safety case conclusions, within the context of a broader skills gap regarding technical literacy. Considering the target audience of the safety case is important to ensure interpretability and adoption of the risk information \cite{fischhoff2011communicating}. This means that different forms of communication for different audiences of the safety case may be necessary, for example technical review, regulatory or public communication. The perceptions and feelings of the audience are important, because this influences how messages are received and affects attitudes regarding threats and intended future actions \cite{spiegelhalter2017risk}. Merely stating that a risk exists is not an effective form of communication because it does not take into account how the audience receives the message, nor how (or if) they will change their behavior as a result \cite{fischhoff2011communicating}. 

Government decision-makers at the executive level do not necessarily have the technical expertise required to understand technical information, unlike technical reviewers. This ‘skills gap’ has been mentioned in the US, UK, Germany and other EU Member States \cite{reuel2024position}, \cite{guha_ai_regulatory}. In the UK, the Trustworthy Autonomous Systems Hub held a Regulators Workshop, inviting UK regulators from a number of different industries to discuss the challenge of regulating AI. A consistent theme was that UK regulators were under-staffed when it came to technical experts and had difficulties recruiting experts with a background in AI and computer science \cite{krook2023trustworthy}. In the US, less than $1\%$ of new AI PhD graduates are choosing to forego lucrative private careers to work for the government \cite{ai_index_report_2024}. 

For government decision-makers therefore, particularly at the executive level, communication must cater to this skills gap. One method is to provide clear visual information that non-technical personnel can understand. If technical terms are included, it is important to explain these terms in plain English, using clear and effective language, and assume that technical information may be misunderstood unless explained effectively. These same precautions should be used for communication with the general public. 

Behavioral economics teaches us that the same risk can be framed in a positive or negative light, depending on how that risk is presented. It is best to use negative framing since positive framing may indirectly increase risk-taking behavior. For example, saying that a risk will not happen 90\% of the time may make that risk seem less likely to the audience. Saying a risk will happen $10\%$ of the time, by contrast, may make it seem more likely to occur. In the medical domain, positive framing of risks of taking a certain drug (there is a 90\% chance of no side-effects) can increase a patient’s likelihood of accepting treatment, thereby indirectly increasing risk-taking behaviour \cite{peters2011informing}. 

As a result, we find that tailoring technical and visual communication to audience needs is critical for effective decision-making and use of risk information. 

\subsection{Communicating confidence for a frontier AI safety case}

Bloomfield and Rushby \cite{bloomfield2022assessing} suggest that the work of an evaluator of a system can be captured as part of a ‘sentencing statement’, which forms part of a broader support structure for decision-makers to deploy (or not) a system. An example sentencing statement could take the form :

“On the basis of this case and an examination of other relevant documentation, I judge the proposed system satisfies the claim that ‘deploying the AI system does not pose unacceptable cyber-risk’ 

\begin{itemize}
    \item “\textit{I believe my judgement of this case is sound and valid because. . .}
    \item \textit{I understand the context and criticality of the decision. . .}
    \item \textit{I understand the system. . .}
    \item \textit{I find a clear thread of reasoning from evidence to claim. . .}
    \item \textit{Evidence provided is sufficient/insufficient for evidence-based decision making}
    \item \textit{I have actively explored doubts. . .}
    \item \textit{I have also identified what evidence would be capable of disproving. . .}
    \item \textit{I have considered and addressed biases and fallacies. . . ”}
\end{itemize}

According to our analysis, the best communication for confidence levels is a multi-factor visual diagram that at a minimum includes information about evidence quality, quality of argumentation, degree of scientific agreement and quantified confidence level. Taking inspiration from the approach of \cite{janzwood2020confident} a multi-factor visual diagram for the assessment of a frontier AI cyber misuse case could look like Figure \ref{fig:fig9} below. This figure may be accompanied by a textual clarification statement to clarify the visual information. A sentencing statement may also be provided. A combination of words, numbers and graphics used collectively, allows for greater communication interpretability on the topic \cite{spiegelhalter2017risk}. 

\begin{figure}[htbp]
  \centering
  \includegraphics[width=0.9\textwidth]{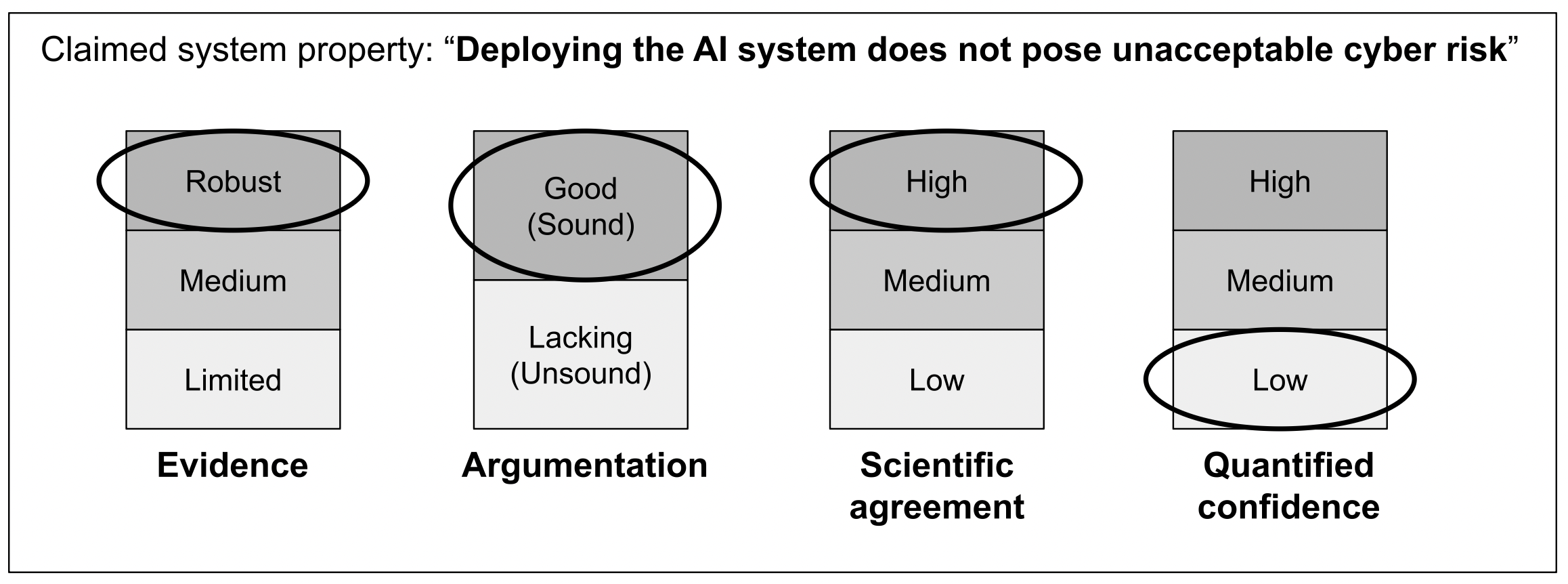}
  \caption{Example visual summary for a safety case for a cyber misuse inability argument.}
  \label{fig:fig9}
\end{figure}

If a safety case has been found to be indefeasible according to Assurance 2.0 then evidence is robust, argumentation is sound and there will be scientific agreement. High ‘scientific agreement’ would be defined to be associated with assessment from a diversely incentivised set of entities, such that a decision-maker can have confidence that the effects of bias have been eradicated. ‘Scientific agreement’ in the context of a cyber misuse safety case would mean that 3rd parties such as regulators or their delegates have been involved with the safety case development, and have thoroughly reviewed it. Quantified confidence is shown low in the example figure above. Such a valuation may very well represent what would be seen if Assurance 2.0 is currently applied in practice to a realistic cyber misuse safety case. One could imagine that it would, justifiably, result in questions from the decision-maker. 

It would also be necessary to provide the decision-maker with clarity on what the words in the top-level claim mean. For example some definition of the ‘system’ should be provided, with accompanying illustration. The term ‘unacceptable risk’ would need defining, for example as follows: ‘\textit{An unacceptable risk is one where there is a probability greater than X\% that sums greater than \$Y billion would be lost, or more than Z in a million lives lost}’. The need for precise and shared meaning of claims increases as the criticality of the system increases. Consideration should be given to providing ontologies that help identify meanings and relate terms to one another, this is important too when common words are being used in a specialist sense.

The gold standard for risk thresholds is a quantitative measurement, however, as we have seen in previous sections this is currently difficult to achieve. Qualitative methods have significant limitations, including significant room for interpretation by practitioners. Therefore, we emphasize that clarity of language and visual language is essential when supplying information that includes qualitative assessments. Risks for example can be effectively communicated with probability distributions, and such visuals are viewed as more desirable than risks expressed numerically or verbally, as the latter can lead to decision-maker bias \cite{duke2024probability}. These curves show the shape of distribution, which gives more information than merely stating “There is a 10\% risk of X.” An example of the type of visualisation which a decision-maker might find useful for the cyber misuse example is shown below in Figure \ref{fig:fig10}, however, the analysis we have discussed in this paper for the cyber misuse safety case would not provide such a figure. Hence this is a topic which could be considered in any further future work.

\begin{figure}[htbp]
  \centering
  \includegraphics[width=0.9\textwidth]{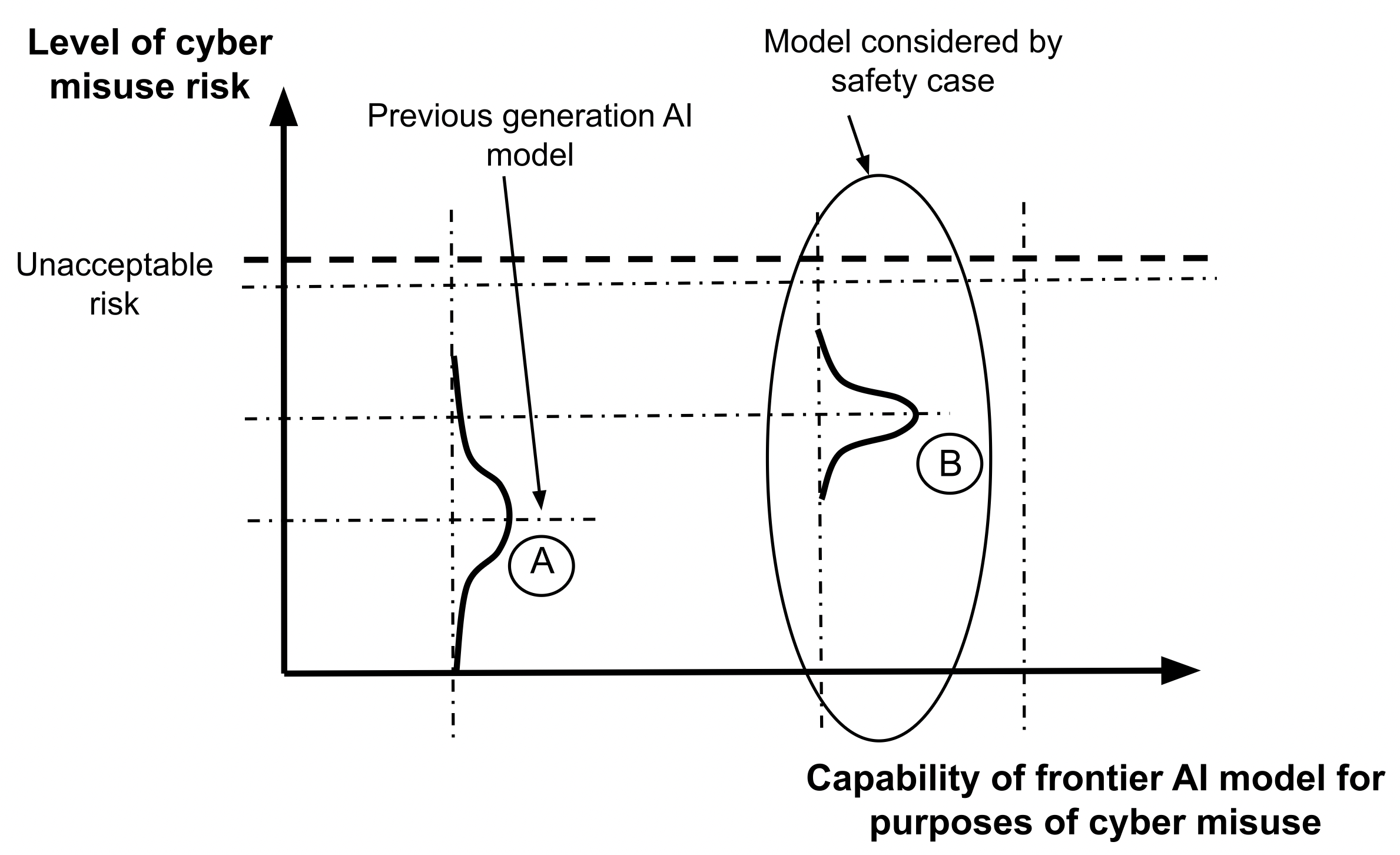}
  \caption{Probability density function based visualisation of risk for the cyber misuse case.}
  \label{fig:fig10}
\end{figure}

For some audiences it may be useful to provide a compressed representation of the safety case, for example a “safety-case on a page”. When doing this, care needs to be taken that the summary remains true to the content of the original argument whilst emphasising the parts that are most important for the particular stakeholder that is using the summary \cite{chozos_miniguide6}. 

\subsection{Testing adequacy of communication}

Visuals used for safety cases need to be tested and evaluated to ensure they communicate effectively their core message to a decision-maker \cite{spiegelhalter2017risk}. By effective, we mean that the decision-maker should be able to understand the visual language, the core message being conveyed and the key figures and statistics, without extrinsic materials. The visual communication is also effective if it helps the decision-maker reach an informed decision on whether to deploy a particular AI system, according to the safety case criteria. 

The U.S. FDA  \cite{fischhoff2011communicating} describes this process of testing visuals as follows:

“A formal evaluation will dispassionately reveal whether a communication is effective. If it’s not, the communication should be revised or discontinued, freeing up resources so that more promising communications get wider audiences. Future work can then build on the evidence of successes and failures”.

A formal evaluation involves a randomized trial with different target groups and iterative stakeholder feedback, focus groups or usability studies \cite{decide2013developing}. To remove bias from this process, the creators of the communication strategy should not be involved in the feedback rounds.

\section{Conclusions}\label{sec10}

\subsection{Findings on confidence assessment}

\begin{enumerate}
    \item A safety case can make a very straight-forward top-level claim, e.g. “Deploying the AI system does not pose unacceptable cyber risk”. However, providing a graduated level of confidence in that simple binary statement is not at all trivial, and indeed requires consideration along multiple dimensions including assessment of logical soundness and completeness, confidence in the mitigation of biases and fallacies and confidence that all reasons for doubt have been elicited and dealt with. Only then might one attempt to additionally try and provide a quantified value of confidence.
    \item We found that when using Assurance 2.0’s ‘product’ or ‘sum of doubts’ methods, in order to achieve even modestly high probabilistic confidence $(95\%)$ in a small safety case fragment having 7 elements (Figure \ref{fig:fig3}), it was required that confidence in each of these argument elements, if evenly distributed, had to be $~99.3\%$. The required confidence in argument elements also only gets higher as the size of the argument under consideration increases. Determining a quantitative value of confidence for leaf nodes will often entail experts making a subjective judgement. This is because data on which confidence might be based, might either not exist or its applicability and relevance to an imperfectly predictable future and post-deployment environment may be uncertain. Achieving extremely high confidence probabilities, under such circumstances seems very challenging. This is particularly true for concerns such as frontier AI cyber misuse where the AI technology that we seek to manage is very novel and the relevant operating environment is vast, and prone to dynamic change.
    \item Because of these factors there seems high likelihood that, in practice, overall confidence in a completed version of the safety case \cite{goemans2024safety} could easily drop very low, even to zero, using the probabilistic valuation methods of Assurance 2.0. Under such circumstances, some of the benefits that one might have hoped to enjoy with probabilistic valuation will not be achieved. Such hoped-for benefits included, i) providing decision-makers with a graduated view about the truthfulness of an otherwise binary claim, ii) providing a determination of when safety assurance case work could be stopped, because adequate confidence has been reached. 
    \item For the above reasons, Bloomfield and Rushby \cite{bloomfield2022assessing} make the point that Assurance 2.0’s quantitative probabilistic valuations of confidence will be of most use for relativistic studies and for ‘what if’ experiments, and not for the purposes of achieving a trusted absolute value of confidence in the overall argument, and based on our studies, we concur. 
    \item It may well be that the achievement of intersubjectivity is the most realistic target, i.e. the development of a very detailed shared understanding, e.g. between AI developers and regulators, on the risks and the rigour that has been applied in trying to minimise those risks. The safety case methodology and the associated confidence assessment techniques will certainly be helpful in this regard. 
    \item In our work, we showed a process for determining probability valuations using a Delphi process, which was operated using LLM ‘experts’. We initially took this approach because we wanted to explore the Delphi approach, but did not have access to, or time to engage actual human experts. However, having conducted the experiment, we think techniques like this may well have applicability in practice, at least in the longer term. It is clear that at a minimum, any viable probabilistic values should be obtained via processes that are both credible and reproducible, especially if they are to be reliably assessed by third parties. Delphi approaches based around the use of LLMs should result in the production of valuations which are reproducible and achieved by mechanisms that are more explicit and transparent. Our preliminary studies indicate that LLMs can predict certain events as well as or better than humans. More work would be needed to confirm whether or not this is robustly true for all pertinent questions and required estimates relevant to the frontier AI / cyber misuse safety case domain, and this will clearly have dependency on at least the capability of the model and the type and the up-to-dateness of the data available during training and inference. Either way, it seems plausible that this may become an approach that could become increasingly useful as the capability of LLMs increases in years to come. In the interim however, we would foresee that human experts would be used in a Delphi process either on their own or augmented with AI/LLMs. 
    \item One challenge we found with the cyber misuse safety case which is likely to also apply to other types of frontier AI related harms, relates to the handling of potential a priori unknown novel harms. The cyber misuse safety case \cite{goemans2024safety} acknowledges the possibility of AI facilitated novel cyberattacks, and proposes to handle this possibility by making use of detection and response mechanisms. However, since the cyber attack mechanism is novel, by definition, it can be expected that there will generally be some residual doubt remaining in a safety case related to whether or not the detection technology will actually be able to detect the novel attack. To determine whether the risk associated with such residual doubt is minor or significant and whether the related residual risk can be accepted, Assurance 2.0 requires a risk assessment to be performed. Two observations can be made. Firstly, it raises the question of how and whether it is possible to adequately perform such a risk assessment given the a priori unknown novel mechanisms involved. The second observation is that Assurance 2.0’s confidence assessment, which is primarily concerned with making an external assessment of the confidence in the argument itself (See Section 2.1), needs to include an aspect of a very different character, being a requirement to make a direct assessment of the top-level internal claim “\textit{C1.1}: Deploying the AI system does not pose unacceptable cyber risk ”. 
    \item Effective communication of confidence to the potentially time-limited and non-expert consumers of this information, such as regulators and executives who may need to make deployment decisions is very challenging. We recommend use of a multi-dimensional visual with textual clarification, outlining at a minimum, information about evidence quality, quality of argumentation, degree of scientific agreement and quantified confidence level, to clarify information along multiple dimensions. 
    \item Safety case development and system development should go hand in hand. Defeaters may be eliminated through system and/or safety case re-design. The order in which defeaters are tackled affects the efficiency with which an AI developer workforce can converge upon an acceptable system design and associated safety case. In Section 8, we described a methodology that AI developers can follow. 
\end{enumerate}

\subsection{Future research possibilities}

\begin{enumerate}
    \item In general, more research is needed on the theoretical foundations of safety case confidence assessment and best practices for implementation. 
    \item There are at least two approaches that might be further considered to address the challenge of probabilistic confidence potentially falling toward zero for a practical and realistic safety case such as that for cyber misuse. Firstly, it can be noted that our results for the cyber misuse safety case fragment (Figure \ref{fig:fig3}) were for a particular argument structure (logically conjunctive and having independent claims). It would be worth considering how things might change, and potentially improve, if and when alternative argument structures are possible, for example where multiple diverse sub-arguments could be used to support a single claim using confidence building blocks. Secondly, in Section 2.1, we discussed two contrasting notions of probabilistic assessment of an argument: i) internal (where a probabilistic claim forms part of the argument itself) and ii) external (where a probabilistic assessment is made about the argument). We observed that a system designer could potentially consider one of two strategies:
    \begin{itemize}
        \item Strategy 1: to adopt a strong claim, for which there might in turn be less confidence in the supporting argument
        \item Strategy 2: adopt a weaker probabilistic claim, but for which there might be greater probabilistic confidence in the argument supporting the claim. 
    \end{itemize}
    Bloomfield and Rushby suggest that where the objective is to make some strongly argued probabilistic assessment of the system, then where possible, the preference should be to include a probabilistic valuation as an internal part of the argument (i.e. along the lines of Strategy 2 above). One reason is that when the probabilistic valuation of confidence is done external to the argument then it uses an argument-generic approach. In contrast, by writing a probabilistic top-level claim, it can focus attention on modifying the context-specific argument itself and/or the associated system in order to achieve the desired probabilistic properties, e.g through mechanisms such as i) use of alternative architectural approaches or ii) use of any theories that may exist that describe how probabilities or context-specific residual doubts may be propagated through a case. In the cyber misuse case, the top-level claim “Deploying the AI system does not pose unacceptable cyber risk”, is supported by an argument wherein the reason for the absence of cyber-risk is due to the inability for harm to arise, either through the inability of the frontier AI to cause harm using conventional types of cyber-attack or through the presence of a control mechanism that can detect and mitigate novel attacks. As such, there is no probabilistic element internal to the argument. However, the top-level claim could alternatively be described in more probabilistic terms. For example, it might be re-written in a form such as ‘The probability of cyber misuse harm exceeding \$X billions, or Y lives, is no more than Z\%’. Consequently, it could be interesting, in any future work, to compare and contrast an approach where there is a probabilistic element internal to the argument with the existing argument where such an element did not exist. 
    \item Whilst methods exist for making a quantitative assessment of the confidence in a top-level claim, the mechanisms available in Assurance 2.0, which include the product method and the sum of doubts method, sometimes only provide approximations or conservative probabilistic valuations. Hence, future work could seek to understand whether improvements might be made to address this issue. 
    \item The robustness of quantitative assessment of confidence of BBNs and DST has been challenged previously \cite{graydon2017investigation}. Bloomfield and Rushby \cite{bloomfield2022assessing} argue that Assurance 2.0 will not suffer from the concerns that were raised. This is also something that would be useful to confirm in any future work.
    \item More work could be conducted into best practices for determining argument leaf node probabilities, which currently often require subjective judgement, and which hence are not necessarily easily reproducible, thereby affecting credibility. It may be useful for example to determine a mapping between types of required probabilistic assessment and the corresponding best practices for obtaining probability values. Such a mapping, if available, might assist practitioners. One such type of probability assessment we investigated was the type that requires some degree of foresight. For this type, we found that frontier AI itself, in the form of LLMs, can enhance traditional foresight methods such as the Delphi method. Going forward, it would be worth exploring how hybrid LLM / human expert approaches could be exploited. 
    \item In making a probabilistic valuation of confidence using Assurance 2.0, we found that there were a number of occasions where threshold values would have to be determined. One such example is the need to select  threshold values for which  weight of evidence could be deemed to be adequate. Another even more fundamental threshold that would have to be determined is that for the overall absolute level of confidence that would be acceptable, where this is assumed to be proportionate to the criticality of the system. Even a probabilistic confidence value like 0.95, which we have noted would likely be very hard to achieve, could be far below what would actually be considered acceptable for high risk systems. Further work could be performed to provide the methods for determining such threshold values. 
    \item Likewise, a method is not provided in Assurance 2.0 for how to conduct the risk assessment that enables the risks associated with residual doubts to be accepted. To assess such risk, the safety engineer will need to take a view on how the residual doubt might propagate through the argument, what the likelihood of occurrence would be and what harm might result. If possible, it would be useful to describe a process that should be followed such that the steps the safety engineer has to take can be made explicit, transparent and less subjective than might otherwise be the case. 
    \item For each specific (mis)use case of frontier AI there will be a need to develop ‘theories’ and associated argument templates. In our work on the hypothetical cyber misuse safety case fragment, we described one such potential theory that might support a substitution block (see Figure \ref{fig:fig4}). However, it could be expected that there will be multiple such theories required in a fully fleshed out safety case. If such theories and their corresponding argument templates are identified as being of a type that might be generally useful to multiple frontier AI developers, or across multiple versions of models, it could be beneficial for them to be standardised in some way. 
    \item The use of natural language in safety cases can cause problems, for example on occasion we had difficulty interpreting the precise meaning of words used in the cyber misuse argument \cite{goemans2024safety}. Future work could include finding or developing a set of templates and terminology for objectively and unambiguously defining evidence, claims, defeaters and arguments applicable for Frontier AI Safety cases. We see benefit in adopting approaches like that seen in requirements capture \footnote{e.g. “EARS = Easy Approach to Requirements Syntax”, see \url{https://www.iaria.org/conferences2013/filesICCGI13/ICCGI_2013_Tutorial_Terzakis.pdf}}.
    \item Our method of prioritizing the order in which defeaters are tackled could be enhanced to cover the additional case where there are dependencies between defeaters as outlined in Section 8.6. 
    \item Further work could consider how safety case confidence probabilities could be updated based on post-deployment safety performance indicator monitoring, see \cite{carlan2024dynamic}.
\end{enumerate}

\subsection{Recommendations}

\begin{enumerate}
    \item Application of techniques like Assurance 2.0 will certainly improve confidence in a safety case through the ensuing systematic consideration of the many important and relevant dimensions to the challenge of confidence assessment. We found that there would undoubtedly be a significant amount of work for frontier AI developers in performing the Assurance 2.0 confidence assessment process. However, as AI capabilities increase leading to an increased risk of harm, such rigour will become increasingly warranted. It was also not evident that short-cuts to this process could or should be made. 
    \item As far as we are aware, there are no standardised guidelines for what methods of confidence assessment should be used by developers of frontier AI. It would therefore be useful to develop and provide such guidelines. Any regulations could then potentially stipulate that the guidelines be followed.
    \item For defeater identification to be comprehensive and credible, it will require involvement of ‘safety-case red-teamers’ who are external to the AI developer.
\end{enumerate}

\section{Acknowledgements}

We would like to express special thanks to Marie Buhl for her help in scoping the initial plan of work, for providing directional guidance throughout the project and for her detailed feedback. We wish to thank Benjamin Hilton for his directional feedback and review comments. We are very grateful also to the creators of Assurance 2.0, Robin Bloomfield and John Rushby for the review comments that they provided. Special thanks too to Kay Kozaronek for his comprehensive and detailed review comments. We wish to thank Adnan Mahmud and Nicola Ding for their review comments. We are grateful to Justin Olive and Ben R Smith of Arcadia Impact for initiating and project managing the AI Governance Taskforce, under which this work has been developed. Alejandro Tlaie and Philip Fox also wish to pass on thanks to Coralie Consigny for the useful discussions they had on forecasting best practices. Thanks too to Anne Le Roux for her comments early in the pre-project phase regarding potentially valuable research directions. 

\bibliographystyle{unsrt}


\begin{thebibliography}{10}

\bibitem{buhl2024safety}
Marie~Davidsen Buhl, Gaurav Sett, Leonie Koessler, Jonas Schuett, and Markus
  Anderljung.
\newblock Safety cases for frontier {AI}.
\newblock {\em arXiv preprint arXiv:2410.21572}, 2024.

\bibitem{bloomfield2022assessing}
Robin Bloomfield and John Rushby.
\newblock Assessing confidence with assurance 2.0.
\newblock {\em arXiv preprint arXiv:2205.04522}, 2022.

\bibitem{goemans2024safety}
Arthur Goemans, Marie~Davidsen Buhl, Jonas Schuett, Tomek Korbak, Jessica Wang,
  Benjamin Hilton, and Geoffrey Irving.
\newblock Safety case template for frontier {AI}: A cyber inability argument.
\newblock {\em arXiv preprint arXiv:2411.08088}, 2024.

\bibitem{clymer2024safety}
Joshua Clymer, Nick Gabrieli, David Krueger, and Thomas Larsen.
\newblock Safety cases: How to justify the safety of advanced {AI} systems.
\newblock {\em arXiv preprint arXiv:2403.10462}, 2024.

\bibitem{anthropic_scaling_policy}
Anthropic's responsible scaling policy.
\newblock Accessed 7 December 2024. Available at:
  \url{https://www.anthropic.com/news/anthropics-responsible-scaling-policy}.

\bibitem{graydon2017investigation}
P.~J. Graydon and C.~M. Holloway.
\newblock An investigation of proposed techniques for quantifying confidence in
  {Assurance} arguments.
\newblock {\em Safety Science}, 92:53--65, 2017.

\bibitem{bloomfield2024confidence}
Robin Bloomfield and John Rushby.
\newblock Confidence in assurance 2.0 cases.
\newblock In {\em The Practice of Formal Methods: Essays in Honour of Cliff
  Jones, Part I}, pages 1--23. Springer, 2024.

\bibitem{diemert2024practitioners}
Simon Diemert, Caleb Shortt, and Jens~H Weber.
\newblock How do practitioners gain confidence in assurance cases?
\newblock {\em arXiv preprint arXiv:2411.03657}, 2024.

\bibitem{groarke1999deductivism}
L.~Groarke.
\newblock Deductivism within pragma-dialectics.
\newblock {\em Argumentation}, 13(1):1--16, 1999.

\bibitem{dalrymple2024towards}
David Dalrymple, Joar Skalse, Yoshua Bengio, Stuart Russell, Max Tegmark,
  Sanjit Seshia, Steve Omohundro, Christian Szegedy, Ben Goldhaber, Nora
  Ammann, et~al.
\newblock Towards guaranteed safe ai: A framework for ensuring robust and
  reliable ai systems.
\newblock {\em arXiv preprint arXiv:2405.06624}, 2024.

\bibitem{bloomfield2020assurance}
Robin Bloomfield and John Rushby.
\newblock Assurance 2.0: A manifesto.
\newblock {\em arXiv preprint arXiv:2004.10474}, 2020.

\bibitem{bloomfield2014building}
R.~E. Bloomfield and K.~Netkachova.
\newblock Building blocks for {Assurance} cases.
\newblock In {\em Proceedings of the 2nd International Workshop on {Assurance}
  Cases for Software-intensive Systems (ASSURE)}. International Symposium on
  Software Reliability Engineering (ISSRE), 2014.

\bibitem{pollock1987defeasible}
J.~L. Pollock.
\newblock Defeasible reasoning.
\newblock {\em Cognitive Science}, 11(4):481--518, 1987.

\bibitem{goodenough2015eliminative}
John~B Goodenough, Charles~B Weinstock, and Ari~Z Klein.
\newblock Eliminative argumentation: A basis for arguing confidence in system
  properties.
\newblock {\em Software Engineering Institute, Carnegie Mellon University,
  Pittsburgh, PA, Tech. Rep. CMU/SEI-2015-TR-005}, 2015.

\bibitem{pegasus_spyware}
Pegasus (spyware).
\newblock 2024. In Wikipedia.
  \url{https://en.wikipedia.org/w/index.php?title=Pegasus_(spyware)&oldid=1261562226}.

\bibitem{stuxnet}
Stuxnet.
\newblock 2024. In Wikipedia.
  \url{https://en.wikipedia.org/w/index.php?title=Stuxnet&oldid=1257264443}.

\bibitem{dalkey1963delphi}
N.~Dalkey and O.~Helmer.
\newblock An experimental application of the {Delphi} method to the use of
  experts.
\newblock {\em Management Science}, 9(3):458--467, 1963.

\bibitem{khodyakov2023rand}
D.~Khodyakov, S.~Grant, J.~Kroger, and M.~Bauman.
\newblock {RAND} methodological guidance for conducting and critically
  appraising delphi panels.
\newblock Technical report, RAND Corporation, 2023.

\bibitem{lewis2020retrieval}
P.~Lewis, E.~Perez, A.~Piktus, F.~Petroni, V.~Karpukhin, N.~Goyal, H.~Küttler,
  et~al.
\newblock Retrieval-augmented generation for knowledge-intensive {NLP} tasks.
\newblock In {\em Advances in Neural Information Processing Systems},
  volume~33, pages 9459--9474. Curran Associates, Inc., 2020.

\bibitem{qin2023toolllm}
Y.~Qin, S.~Liang, Y.~Ye, K.~Zhu, L.~Yan, Y.~Lu, Y.~Lin, et~al.
\newblock {ToolLLM}: Facilitating {Large Language Models} to master 16000+
  real-world {APIs}, 2023.

\bibitem{harney2003bayesian}
H.~L. Harney.
\newblock {\em Bayesian Inference}.
\newblock Springer, Berlin, Heidelberg, 2003.

\bibitem{kunda1990motivated}
Z.~Kunda.
\newblock The case for motivated reasoning.
\newblock {\em Psychological Bulletin}, 108(3):480--498, 1990.

\bibitem{nickerson1998confirmation}
Raymond~S Nickerson.
\newblock Confirmation bias: A ubiquitous phenomenon in many guises.
\newblock {\em Review of general psychology}, 2(2):175--220, 1998.

\bibitem{wason1960failure}
P.~C. Wason.
\newblock On the failure to eliminate hypotheses in a conceptual task.
\newblock {\em Quarterly Journal of Experimental Psychology}, 12(3):129--140,
  1960.

\bibitem{dunning1990overconfidence}
D.~Dunning, D.~W. Griffin, J.~D. Milojkovic, and L.~Ross.
\newblock The overconfidence effect in social prediction.
\newblock {\em Journal of Personality and Social Psychology}, 58(4):568--581,
  1990.

\bibitem{tversky1974judgment}
A.~Tversky and D.~Kahneman.
\newblock Judgment under uncertainty: Heuristics and biases.
\newblock {\em Science}, 185(4157):1124--1131, 1974.

\bibitem{tversky1973availability}
A.~Tversky and D.~Kahneman.
\newblock Availability: A heuristic for judging frequency and probability.
\newblock {\em Cognitive Psychology}, 5(2):207--232, 1973.

\bibitem{janis1972victims}
I.~L. Janis.
\newblock {\em Victims of Groupthink: A Psychological Study of Foreign-Policy
  Decisions and Fiascoes}.
\newblock Houghton Mifflin, Oxford, England, 1972.

\bibitem{milgram1963behavioral}
S.~Milgram.
\newblock Behavioral study of obedience.
\newblock {\em The Journal of Abnormal and Social Psychology}, 67(4):371--378,
  1963.

\bibitem{gohar2024codefeater}
U.~Gohar, M.~C. Hunter, R.~R. Lutz, and M.~B. Cohen.
\newblock Codefeater: Using {LLMs} to find defeaters in {Assurance} cases.
\newblock In {\em Proceedings of the 39th IEEE/ACM International Conference on
  Automated Software Engineering (ASE '24)}, pages 2262--2267, New York, NY,
  USA, 2024. Association for Computing Machinery.

\bibitem{leveson2011safety}
N.~G. Leveson.
\newblock The use of safety cases in certification and regulation.
\newblock Technical report, Massachusetts Institute of Technology, Engineering
  Systems Division, 2011.
\newblock Working Paper.

\bibitem{uuk2024effective}
R.~Uuk, A.~Brouwer, N.~Dreksler, V.~Pulignano, and R.~Bommasani.
\newblock Effective mitigations for systemic risks from general-purpose {AI},
  2024.
\newblock SSRN Scholarly Paper. Rochester, NY: Social Science Research Network.

\bibitem{carlan2024dynamic}
C.~Cârlan, F.~Gomez, Y.~Mathew, K.~Krishna, R.~King, P.~Gebauer, and B.~R.
  Smith.
\newblock Dynamic safety cases for frontier {AI}, 2024.

\bibitem{varadarajan2023clarissa}
S.~Varadarajan, R.~Bloomfield, J.~Rushby, G.~Gupta, A.~Murugesan, R.~Stroud,
  K.~Netkachova, and I.~H. Wong.
\newblock {CLARISSA}: Foundations, tools \& automation for {Assurance} cases.
\newblock In {\em 2023 IEEE/AIAA 42nd Digital Avionics Systems Conference
  (DASC)}, pages 1--10, Barcelona, Spain, 2023. IEEE.

\bibitem{eppler2009systematic}
M.~J. Eppler and M.~Aeschimann.
\newblock A systematic framework for risk visualization in risk management and
  communication.
\newblock {\em Risk Management}, 11(2):67--89, 2009.

\bibitem{trevena2013presenting}
L.~J. Trevena, B.~J. Zikmund-Fisher, A.~Edwards, W.~Gaissmaier, M.~Galesic,
  P.~K.~J. Han, J.~King, et~al.
\newblock Presenting quantitative information about decision outcomes: A risk
  communication primer for patient decision aid developers.
\newblock {\em BMC Medical Informatics and Decision Making}, 13(Suppl 2):S7,
  2013.

\bibitem{hausmann2008sequential}
D.~Hausmann and D.~Läge.
\newblock Sequential evidence accumulation in decision making: The individual
  desired level of confidence can explain the extent of information
  acquisition.
\newblock {\em Judgment and Decision Making}, 3(3):229--243, 2008.

\bibitem{spiegelhalter2017risk}
D.~Spiegelhalter.
\newblock Risk and uncertainty communication.
\newblock {\em Annual Review of Statistics and Its Application}, 4(1):31--60,
  2017.

\bibitem{fischhoff2011communicating}
B.~Fischhoff, N.~T. Brewer, and J.~S. Downs.
\newblock Communicating risks and benefits: An evidence-based user's guide.
\newblock Technical report, FDA, 2011.

\bibitem{reuel2024position}
A.~Reuel, L.~Soder, B.~Bucknall, and T.~A. Undheim.
\newblock Position paper: Technical research and talent is needed for effective
  {AI} governance, 2024.

\bibitem{guha_ai_regulatory}
N.~Guha, C.~M. Lawrence, L.~A. Gailmard, K.~T. Rodolfa, F.~Surani,
  R.~Bommasani, I.~D. Raji, et~al.
\newblock The {AI} regulatory alignment problem.
\newblock \url{https://dho.stanford.edu/wp-content/uploads/AI_Regulation.pdf}.

\bibitem{krook2023trustworthy}
J.~Krook et~al.
\newblock Trustworthy autonomous systems hub -- written evidence.
\newblock Technical report, House of Lords Communications and Digital Select
  Committee, 2023.
\newblock Inquiry: Large language models.

\bibitem{ai_index_report_2024}
{AI} index report 2024 -- artificial intelligence index.
\newblock Accessed 20 December 2024. Available at:
  \url{https://aiindex.stanford.edu/report/}.

\bibitem{peters2011informing}
E.~Peters, P.~S. Hart, and L.~Fraenkel.
\newblock Informing patients: The influence of numeracy, framing, and format of
  side effect information on risk perceptions.
\newblock {\em Medical Decision Making}, 31(3):432--436, 2011.

\bibitem{janzwood2020confident}
S.~Janzwood.
\newblock Confident, likely, or both? the implementation of the uncertainty
  language framework in {IPCC} special reports.
\newblock {\em Climatic Change}, 162(3):1655--1675, 2020.

\bibitem{duke2024probability}
M.~C. Duke.
\newblock Probability and confidence: How to improve communication of
  uncertainty about uncertainty in intelligence analysis.
\newblock {\em Journal of Behavioral Decision Making}, 37(1):e2364, 2024.

\bibitem{chozos_miniguide6}
N.~Chozos and R.~Bloomfield.
\newblock Mini-guide 6: Summarising and communication.
\newblock Part of Declare CAE guidance document set.

\bibitem{decide2013developing}
DECIDE Consortium, S.~Treweek, A.~D. Oxman, P.~Alderson, P.~M. Bossuyt,
  L.~Brandt, J.~Brożek, et~al.
\newblock Developing and evaluating communication strategies to support
  informed decisions and practice based on evidence ({DECIDE}): Protocol and
  preliminary results.
\newblock {\em Implementation Science}, 8(1):6, 2013.

\end{thebibliography}

\end{document}